\documentclass[reprint,twocolumn,notitlepage,preprintnumbers,nofootinbib,amsmath,amssymb,aps,prd,floatfix,superscriptaddress,longbibliography,svgnames]{revtex4-1}

\usepackage{placeins}
\usepackage{dcolumn}
\usepackage{multirow}
\usepackage{subfigure}
\usepackage{physics}
\usepackage{cancel}
\usepackage{stackrel}
\usepackage{paralist}
\usepackage{xspace}
\usepackage{slashed}
\usepackage{cancel}
\usepackage{todonotes}
\usepackage{enumerate}
\usepackage{float}
\usepackage{fullpage}
\usepackage{ulem}
\usepackage{wasysym}
\usepackage{comment}
\usepackage{bbold}
\usepackage{mathtools}
\usepackage{ulem}
\usepackage{booktabs}
\usepackage{multirow}
\usepackage{soul}
\usepackage[breaklinks,colorlinks,urlcolor=blue,citecolor=blue,linkcolor=magenta]{hyperref}
\usepackage{xcolor}
\usepackage{orcidlink}
\usepackage{datetime}

\usepackage{url}
\usepackage{enumitem}
\usepackage{graphicx}

\newcommand{\lcdm }{$\Lambda$CDM }

\usepackage{feynmp-auto}
\newcommand{\nua}[1]{\ensuremath{\rlap{\kern-2.5pt\ensuremath{\overset{\scriptscriptstyle(-)}{\phantom{\nu}}}}{\ensuremath{{\nu}_{#1}}}}}

\newcommand{\summnu}{\sum \tilde{m}_\nu}

\begin{document}

\title{A short blanket for cosmology:\\ the CMB lensing anomaly behind the preference for a negative neutrino mass}

\author{Andrea Cozzumbo\orcidlink{0009-0004-2772-2692}}
\email{andrea.cozzumbo@gssi.it}
\affiliation{Gran Sasso Science Institute (GSSI), Viale F. Crispi 7, L'Aquila (AQ), I-67100, Italy}
\affiliation{INFN - Laboratori Nazionali del Gran Sasso (LNGS), L'Aquila (AQ), I-67100, Italy}

\author{Mattia Atzori Corona \orcidlink{https://orcid.org/0000-0001-5092-3602}}
\email{mcorona@roma2.infn.it}
\affiliation{Istituto Nazionale di Fisica Nucleare (INFN), Sezione di Roma Tor Vergata,
Via della Ricerca Scientifica, I-00133 Rome, Italy}

\author{Riccardo Murgia\orcidlink{0000-0002-2224-7704}}
\email{riccardo.murgia89@unica.it}
\affiliation{Dipartimento di Fisica, Universit\`a degli Studi di Cagliari, Cittadella Universitaria, 09042 Monserrato (CA), Italy}
\affiliation{INFN, Sezione di Cagliari, Cittadella Universitaria, 09042 Monserrato (CA), Italy}
\affiliation{INAF - Osservatorio Astronomico di Cagliari, Via della Scienza 5, 09047 Selargius (CA), Italy}

\author{Maria Archidiacono\orcidlink{0000-0003-4952-9012}}
\email{maria.archidiacono@unimi.it}
\affiliation{Dipartimento di Fisica ``Aldo Pontremoli'', Università degli Studi
di Milano, Via Celoria 16, 20133 Milano, Italy}
\affiliation{INFN, Sezione di Milano, Via Celoria 16, 20133 Milano, Italy}

\author{Matteo Cadeddu\orcidlink{0000-0002-3974-1995}}
\email{matteo.cadeddu@ca.infn.it}
\affiliation{INFN, Sezione di Cagliari, Cittadella Universitaria, 09042 Monserrato (CA), Italy}


\begin{abstract}
Recent analyses combining cosmic microwave background (CMB) and baryon acoustic oscillation (BAO) challenge particle physics constraints on the total neutrino mass, pointing to values smaller than the lower limit from neutrino oscillation experiments. To examine the impact of different CMB likelihoods from \textit{Planck}, lensing potential measurements from \textit{Planck} and ACT, and BAO data from DESI, we introduce an effective neutrino mass parameter ($\summnu$) which is allowed to take negative values.
We investigate its correlation with
two extra parameters capturing the impact of gravitational lensing on the CMB: one controlling the smoothing of the peaks of the temperature and polarization power spectra; one rescaling the lensing potential amplitude. In this configuration, we infer $\summnu=-0.018^{+0.085}_{-0.089}~\text{eV}~(68\% ~\text{C.L.})$, which is fully consistent with the minimal value required by neutrino oscillation experiments.
We attribute the apparent preference for negative neutrino masses to an excess of gravitational lensing detected by late-time cosmological probes compared to that inferred from \textit{Planck} CMB angular power spectra. We discuss implications in light of the DESI BAO measurements and the CMB lensing anomaly.
\end{abstract}

\maketitle  

\section{Introduction}\label{sec:intro}

Several recent studies have performed joint analyses of cosmic microwave background (CMB) anisotropies from \textit{Planck} and baryon acoustic oscillation (BAO) measurements from the Dark Energy Spectroscopic Instrument (DESI) to constrain the sum of neutrino masses, $\sum m_\nu$~\cite{DESI:2025ejh,Craig:2024tky,Green:2024xbb,Loverde:2024nfi,Lynch:2025ine,Naredo-Tuero:2024sgf,Reboucas:2024smm,Jiang:2024viw, Elbers:2024sha, Sailer:2025lxj, Jhaveri:2025neg, Herold:2024nvk, Noriega:2024lzo, Chebat:2025kes, Giare:2025ath, Namikawa:2025doa, Chen:2025mlf,Du:2025xes, GarciaEscudero:2025lef, Allali:2025yvp, Graham:2025fdt, Sharma:2025iux}. A particularly striking outcome from these studies is that, when negative values of $\sum m_\nu$ are allowed within the $\Lambda$CDM framework, the combined analysis exclude the massless case at approximately $2\sigma$~\cite{DESI:2025ejh, Elbers:2024sha, Green:2024xbb, Craig:2024tky}. This apparent preference for negative values reflects the degeneracy between $\sum m_\nu$ and parameters controlling the late-time expansion rate, since variations in $\sum m_\nu$ alter the comoving sound horizon and hence the angular scales measured by CMB and BAO observations~\cite{Sharma:2025iux}.

In the standard \lcdm framework, the ratio of the angular scale of the sound horizon at recombination ($z_*$) and at the redshift of BAO ($z_\mathrm{BAO}$), $\theta_s / \theta_\mathrm{BAO}$, is expected to remain stable -- unless modifications to the expansion history occur in the post-recombination era~\cite{Bernal:2020vbb, Loverde:2024nfi, Knox:2019rjx}. 
However, the aforementioned analyses have revealed a mild yet persistent discrepancy between CMB and BAO geometrical measurements. This mismatch can be interpreted as a tension in the value of total matter abundance $\Omega_\mathrm{m}$ inferred at the two epochs. CMB data favors a higher matter density compared to that inferred from BAO, with a significance of about $2.3\sigma$~\cite{Planck:2018vyg,DESI:2025zgx}. This issue is commonly referred to as the ``matter deficit" problem~\cite{Lynch:2025ine, Loverde:2024nfi}.
The neutrino mass anomaly emerges as the fit attempts to resolve this issue pushing $\sum m_\nu$ to negative values to reduce the late-time total matter abundance.
In terms of linear perturbations, a negative value of $\sum m_\nu$ would instead boost gravitational clustering at late time, enhancing the small-scale linear matter power spectrum and the gravitational lensing of CMB.

Since different physical effects can produce similar observational signatures, the cosmological preference for $\sum m_\nu < 0$ can be used as a proxy to investigate potential underlying inconsistencies among complementary data sets.
To address this issue, it is essential to examine degeneracies with other cosmological parameters. One example is the correlation between $\sum m_\nu$ and the optical depth to reionization, $\tau_\mathrm{reio}$~\cite{Jhaveri:2025neg, Sailer:2025lxj, Green:2024xbb, Archidiacono:2016lnv}. The small-scale CMB anisotropies are primarily sensitive to the parameter combination $A_s e^{-2\tau_\mathrm{reio}}$, where $A_s$ denotes the amplitude of primordial scalar perturbations. Increasing $A_s$ enhances the gravitational lensing potential, thus compensating for the opposite effect of massive neutrinos. In turn, the higher value of $A_s$ needs to be compensated by a higher value of $\tau_\mathrm{reio}$ in order to fit the overall amplitude of the temperature power spectrum, leading to a degeneracy with $\sum m_\nu$, that can be broken by measuring the reionization bump in low multipole ($\ell$) CMB polarization data. This measurement remains challenging due to its faintness and foregrounds. Different low-$\ell$ likelihoods, such as \texttt{SimAll}~\cite{Planck:2019nip}, \texttt{LoLLiPoP}~\cite{Tristram:2021tvh} and \texttt{SRoll2}~\cite{Pagano:2019tci} incorporate different treatments of these systematics, in order to estimate $\tau_\mathrm{reio}$ from \textit{Planck} polarization data.

Another anomaly arises in the context of \textit{Planck} CMB data at high multipoles, commonly referred to as the ``CMB lensing anomaly”~\cite{Calabrese:2008rt, Planck:2016tof, Planck:2018vyg}. The observed CMB power spectra show an unexpected oscillatory residual pattern at high-$\ell$, relative to standard \lcdm predictions. This anomaly is typically cast in terms of an extra-parameter, $A_\mathrm{lens}$~\cite{Calabrese:2008rt}, which rescales the overall amplitude of CMB lensing such that $A_\mathrm{lens}=1$ is the expected result.
The reported statistical significance of the lensing anomaly depends on the choice of the high-$\ell$ likelihood used in the analysis~\cite{Addison:2023fqc, Naredo-Tuero:2024sgf}. Within $\Lambda$CDM, the 2018 \texttt{Plik} likelihood~\cite{Planck:2018vyg} indicates a discrepancy at the $2.8\sigma$ level. More recent likelihoods, benefiting from improved data processing and better control over systematics, show a lower level of tension: approximately $1.7\sigma$ with \texttt{CamSpec}~\cite{Efstathiou:2019mdh, Rosenberg:2022sdy} and only $0.7\sigma$ with \texttt{HiLLiPoP}~\cite{Tristram:2021tvh, Tristram:2023haj}. 

Gravitational lensing impacts CMB in two ways: through the smoothing of acoustic peaks in the temperature and polarization power spectra, and by scaling the lensing potential $C^{\phi\phi}_\ell$ derived from the CMB trispectrum~\cite{Green:2024xbb, Hu:2001kj, Okamoto:2003zw, Hu:2001fa, ACT:2023dou}. 
To distinguish the two effects, in this work we will adopt a two-parameter approach~\cite{Simard:2017xtw, Murgia:2020ryi, Corona:2021qxl}. We will introduce the parameter $A^{\mathrm{TTTEEE}}_{\mathrm{lens}}$, which controls the smoothing in the high-$\ell$ TT, TE, EE power spectra, along with the parameter $A^{\phi\phi}_{\mathrm{lens}}$, which scales the global amplitude of the lensing potential. 
In absence of systematics, both parameters are expected to equal unity, reflecting consistent measurements from the two methods. \textit{Planck} data seem to point to a value of $A_{\mathrm{lens}}^{\mathrm{TTTEEE}} > 1$,~$i.e.$~to an excess of smoothing of the CMB peaks~\cite{Calabrese:2008rt, Planck:2018vyg, Murgia:2020ryi, Addison:2023fqc}.
Varying the total neutrino mass affects both the smoothing of the acoustic peaks and the amplitude of the lensing potential, so that the impact of a free effective neutrino mass parameter is degenerate with that of $A_{\mathrm{lens}}$.

In this work, we will exploit the two-parameter decomposition outlined above to clarify the connection between the CMB lensing anomaly and the cosmological preference for negative neutrino masses.
We will extensively study the interplay between the lensing amplitude parameters and an effective neutrino mass parameter, $\summnu$, which is allowed to take negative values.
We will assess the impact on the negative neutrino mass and lensing anomalies
of different CMB likelihoods from \textit{Planck}, lensing potential measurements from \textit{Planck} and the
Atacama Cosmology Telescope (ACT), and DESI BAO data~\cite{DESI:2025zgx, DESI:2025ejh}.
We will show that the neutrino mass anomaly is pointing
to an excess of gravitational lensing detected by
late-time cosmological probes (BAO + CMB lensing) compared to that inferred from the smoothing of \textit{Planck} CMB angular power spectra.

This paper is structured as follows: in Section~\ref{sec:data} we describe the data sets and methodology used in our analyses; in Section~\ref{sec:results} we present and discuss our main results; in Section~\ref{sec:conclusion} we provide summary and conclusion.

\section{Data and Methods}\label{sec:data}

We model the phenomenological impact of negative neutrino masses by performing an expansion of all cosmological observables ($X$), both background and linear perturbation level, around their value in the massless neutrino limit ($X^0$)~\cite{Elbers:2024sha}:
\begin{equation}\label{eq:mnu}
    X^{\scriptstyle{\sum} \tilde m_{\nu}} = X^{0} + \mathrm{sgn}\left(\small{\sum} \tilde{m}_{\nu}\right) \left[X^{|\scriptstyle{\sum}  \tilde m_{\nu}|}- X^{0}\right],
\end{equation} 
where $X^{\scriptstyle{\sum} \tilde m_{\nu}}$ is the value of a given observable when the effective mass parameter $\summnu$ is allowed to vary to negative values.
We instead indicate the total actual neutrino mass as $\sum m_\nu \equiv \sum^3_{i=1} m_{\nu,i}$, assuming three degenerate-mass neutrino species.  
We implement Eq.~\eqref{eq:mnu} in the cosmological sampler \texttt{Cobaya}~\cite{Torrado:2020dgo}, that we interface with the numerical Einstein-Boltzmann solver \texttt{CLASS}~\cite{Lesgourgues2011CosmicI, Blas2011Cosmic}. 

We perform extensive Monte Carlo Markov Chain (MCMC) analyses against different combinations of the following data sets:
\begin{itemize}

  \item \texttt{Planck Commander}: the default low-$\ell$ TT CMB likelihood from the \textit{Planck} 2018 (PR3) release, covering the multipole range $2 \leq \ell < 30$ \cite{Planck:2019nip}.

  \item \texttt{Planck SimAll} or \texttt{Sim}: the default low-$\ell$ EE CMB likelihood from the PR3 data release~\cite{Planck:2019nip}. In our analysis, we use the multipole range $2 \leq \ell < 30$.
  
  \item \texttt{Planck SRoll2} or \texttt{SR2}: an improved version of the default low-$\ell$ EE CMB likelihood that corrects for instrumental systematics and a better mapmaking algorithm, which better handles systematics and noise.  In our analysis, we use the multipole range $2 \leq \ell < 30$~\cite{Pagano:2019tci, Delouis:2019bub}.
  
  \item \texttt{Planck LoLLiPoP} or \texttt{LoL}: an alternative likelihood for EE, EB and BB low-$\ell$ spectra in the range $2 \leq \ell < 30$~\cite{Hamimeche:2008ai, Tristram:2021tvh, Mangilli:2015xya} based on the 2020
  NPIPE (PR4) data release. To ensure a fair comparison with the \texttt{SRoll2} likelihood, we perform analyses both with (\texttt{EE}) and without including EB and BB (\texttt{EEEBBB}) spectra.

  \item \texttt{Planck Plik} or \texttt{Plk}: the default high-$\ell$ CMB likelihood from the \textit{Planck} 2018 (PR3) legacy release. We use multipoles in the range $30 \leq \ell \leq 2500$ for TT and $30 \leq \ell \leq 2000$ for TE and EE. We use the TTTEEE configuration, which combines TT, TE, and EE data~\cite{Planck:2018vyg, Planck:2019nip}.
  
  \item \texttt{Planck CamSpec} or \texttt{CSp}: a high-$\ell$ CMB likelihood based on the NPIPE (PR4) data release, incorporating TT, TE, and EE power spectra. Compared to the earlier \texttt{Plik} likelihood, this likelihood benefits from improved detector calibration and more comprehensive systematic corrections. The reduced noise level in the NPIPE maps enhances parameter precision, yielding roughly a 10\% tighter constraint on most $\Lambda$CDM parameters in TTTEEE, driven largely by improvements in polarization data quality~\cite{Efstathiou:2019mdh, Rosenberg:2022sdy}. We use the same multipole ranges as for \texttt{Plik}.
  
  \item \texttt{Planck HiLLiPoP} or \texttt{HiL}: a refined CMB likelihood designed for high-$\ell$ TT, EE, and TE spectra, based on the NPIPE (PR4). It incorporates a larger sky coverage and a wider range of multipoles, leading to more precise cosmological parameter constraints~\cite{Tristram:2021tvh, Tristram:2023haj}. We use the same multipole ranges as for \texttt{Plik}.
  
  \item \texttt{P-ACT lensing}: the joint CMB lensing potential measurements from \textit{Planck} PR4 NPIPE maps and ACT. The complementary information provided by ACT in addition to \textit{Planck}, providing a better overall constraining and better control of systematics~\cite{ACT:2023dou}.

  \item \texttt{DESI}: the DR2 data release from the DESI collaboration~\cite{DESI:2025ejh} including the comoving and line-of-sight distances and relative uncertainties presented in Table IV of \cite{DESI:2025zgx}.

\end{itemize}
\begin{figure*}[t]
    \centering
    \includegraphics[width=\textwidth]{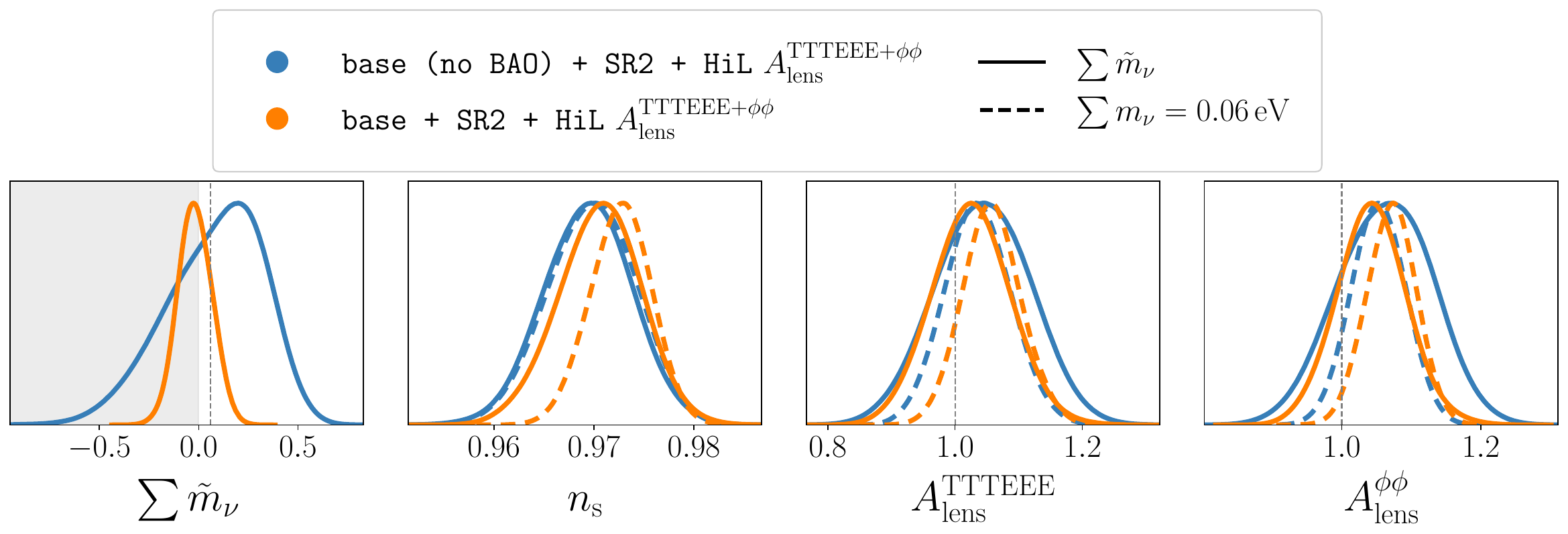}
    \caption{1D marginalized posterior distribution for $\sum \tilde{m}_\nu$, $n_s$, $A^{\mathrm{TTTEEE}}_\text{lens}$ an $A^{\phi\phi}_\text{lens}$. All analyses include \texttt{P-ACT} lensing, low-$\ell$ TT data from \texttt{Commander}, low-$\ell$ EE data from \texttt{SRoll2}, and high-$\ell$ TTTEEE data from \texttt{HiLLiPoP}. Lensing amplitude parameters are varied in all cases. Orange curves include DESI DR2 BAO; blue exclude it. 
    Dashed lines indicate the scenario with fixed neutrino mass, $\sum m_\nu = 0.06~\mathrm{eV}$, while solid lines represent cases where it is allowed to vary freely.
    Vertical dashed lines mark the NO mass limit and the fiducial lensing values. The grey shaded region indicates the unphysical parameter space.}
    \label{fig:alens_1d_sroll}
\end{figure*}

We refer to~\cite{Jense:2025wyg} for an extensive description of differences among various likelihoods.
We will make extensive use of the \texttt{Commander + P-ACT + DESI } data set combination, which we refer to as \texttt{base} for brevity.
We consider the MCMC to have converged once the Gelman-Rubin criterion satisfies $R - 1 < 0.03$~\cite{Gelman:1992zz}.
We carry out our analyses varying the six standard parameters of the \lcdm model: the physical cold dark matter density ($\omega_\mathrm{c}$), the physical baryon density ($\omega_\mathrm{b}$), the angular size of the sound horizon at recombination ($\theta_s$), the amplitude ($A_s$, varied as $\log (10^{10} A_s)$ ) and the spectral index ($n_s$) of the primordial scalar perturbations, and the optical depth to reionization ($\tau_\mathrm{reio}$). In our \lcdm analyses we fix the neutrino mass to the minimal value allowed by neutrino oscillation experiments, assuming normal neutrino mass ordering (NO) $\sum m_\nu = 0.06~\mathrm{eV}$~\cite{ParticleDataGroup:2024cfk}.
We adopt broad, uniform priors on all cosmological parameters and incorporate the full set of recommended nuisance parameters specific to each data set~\cite{DESI:2025ejh}. We also make use of some extra parameters beyond $\Lambda$CDM, particularly the total effective neutrino mass $\summnu$.
Moreover, we introduce two additional parameters: $A^{\mathrm{TTTEEE}}_{\mathrm{lens}}$ and $A^{\phi\phi}_{\mathrm{lens}}$, designed to marginalize over the lensing information in the \textit{Planck} data~\cite{Simard:2017xtw,Murgia:2020ryi,Corona:2021qxl}. The parameter $A^{\phi\phi}_{\mathrm{lens}}$ rescales the amplitude of the lensing potential power spectrum, while $A^{\mathrm{TTTEEE}}_{\mathrm{lens}}$ rescales the amplitude of the acoustic peak smoothing in the CMB temperature and polarization power spectra. We introduce them in our \texttt{CLASS} version (see Appendix~\ref{ap:app-Alens} for details on the implementation).
We obtain best-fit estimates from our MCMC chains with the minimizer \texttt{Py-BOBYQA}~\cite{Cartis:2018xum} within \texttt{Cobaya}.
Given the slightly asymmetric shape of the 1D marginalized posterior distributions, the most appropriate measure of data set compatibility is the fraction of the distribution lying below a chosen reference value, quantified through the Cumulative Distribution Function (CDF)~\cite{Loverde:2024nfi}. We convert this probability into a Gaussian-equivalent significance level by mapping the CDF to a $z$-score using the percent point function. This yields an estimate of the departure from the reference model ($e.g.$, $\sum m_\nu = 0.06~\mathrm{eV}$) in terms of the standard deviation $\sigma$. Hereafter, the shaded MCMC contours in the plots represent the $68\%$ and $95\%$ confidence level (C.L.).

\section{Results and Discussion}\label{sec:results}

Our goal is to determine a consistent combination of CMB likelihoods, data sets and parameter combinations, to relax the negative neutrino mass tension without compromising the overall goodness of the fit. By doing so, we aim to shed light on whether this anomaly might be related to specific modeling assumptions, statistical artifacts or unknown systematics in the data.

Our strategy builds upon the analysis of~\cite{Green:2024xbb} introducing two major improvements: $(i)$ we extend the parameter space to include the impact of negative neutrino masses at both the background and linear perturbation levels, as described in Section~\ref{sec:data}, following~\cite{Elbers:2024sha}; $(ii)$ we select the \texttt{SRoll2 + HiLLiPoP} likelihood combination as our reference CMB data set~\cite{Naredo-Tuero:2024sgf, Allali:2024aiv}. Using \texttt{HiLLiPoP} mitigates the lensing anomaly in the temperature and polarization spectra. The use of \texttt{SRoll2} for low-$\ell$ polarization improves the treatment of large-scale polarization data~\cite{Pagano:2019tci, Delouis:2019bub}, yielding tighter constraints on $\tau_\mathrm{reio}$ and reducing deviations from $\sum m_\nu = 0.06~\mathrm{eV}$ compared to results obtained with \texttt{SimAll}~\cite{DESI:2025gwf}. As shown in Appendix~\ref{ap:app-planck} -- where we provide a comprehensive comparison among different CMB likelihood choices -- the use of \texttt{SRoll2} or \texttt{LoLLiPoP} yields no significant differences.
\begin{figure*}[t]
    \centering
    \includegraphics[width=\textwidth]{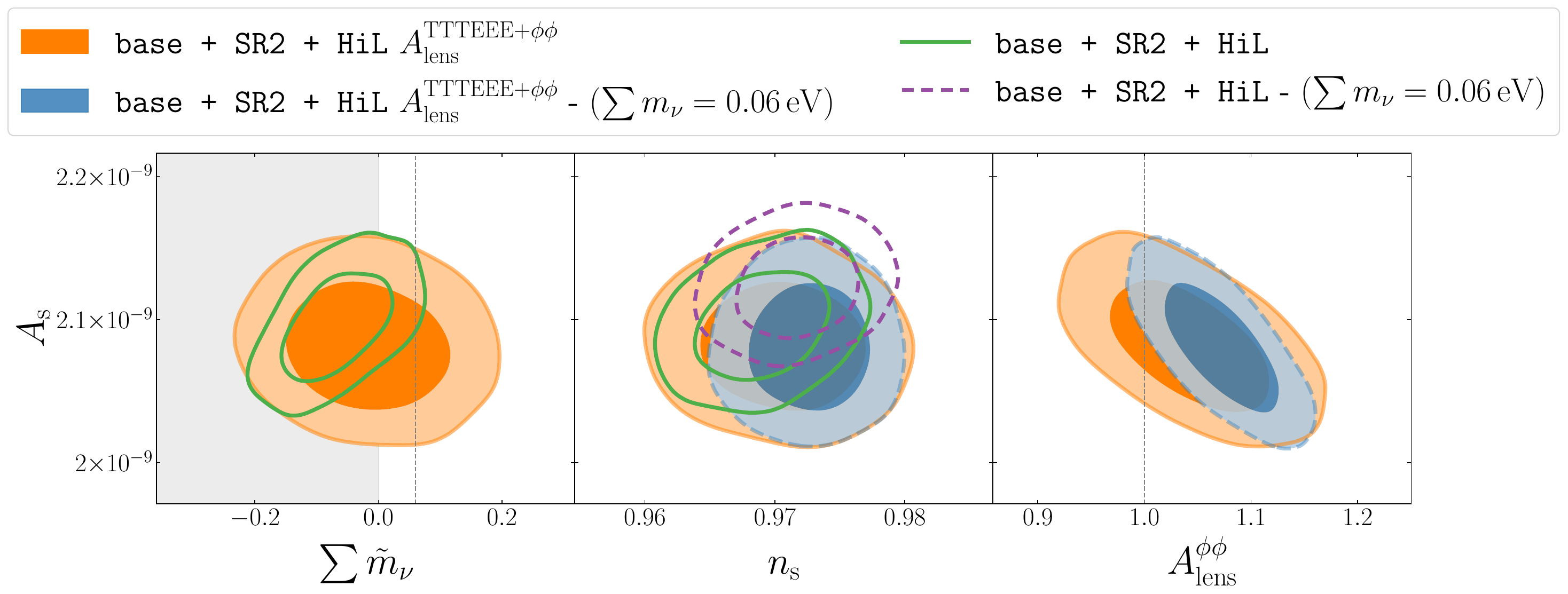}
    \caption{Constraints for $A_s$, $\summnu$, $n_s$ and $A^{\phi\phi}_\text{lens}$. All analyses include the \texttt{base} data sets (DESI DR2 BAO, \texttt{P-ACT} lensing, low-$\ell$ TT from \texttt{Commander}) with the addition of \texttt{SRoll2} and \texttt{HiLLiPoP}. Solid lines denote cases with free neutrino mass, while dashed lines keep the mass fixed to $\sum m_\nu = 0.06~\mathrm{eV}$. Filled contours indicate that the lensing amplitude parameters are varied, empty contours correspond to analyses where they are held fixed.
    Vertical grey dashed lines mark the NO mass limit and the fiducial value of unity for the lensing amplitude parameters. The grey shaded region indicates the unphysical parameter space.}
    \label{fig:alens_2d_sroll}
\end{figure*}
As reported in Section~\ref{sec:intro}, replacing \texttt{Plik} with \texttt{HiLLiPoP} largely reduces the lensing anomaly~\cite{Addison:2023fqc}.
This is because \texttt{HiLLiPoP} resolves the anomaly associated with the smearing of the CMB peaks, \textit{i.e.}, $A^{\mathrm{TTTEEE}}_\text{lens}$ is consistent with unity with this likelihood choice. However, $A^{\phi\phi}_\text{lens}$ still shows a non-negligible, although mild deviation from unity.
Moreover, independently of the choice of the high-$\ell$ TTTEEE likelihood, including \texttt{P-ACT} and \texttt{DESI} data further pushes $A^{\phi\phi}_\text{lens}$ to values larger than 1.
For further details see Appendix~\ref{ap:app-lensing}. 

In Section~\ref{subsec-alens} we will examine the interplay between $\summnu$ with $n_s$, $A_s$, $A^{\mathrm{TTTEEE}}_{\mathrm{lens}}$ and $A^{\phi\phi}_{\mathrm{lens}}$. In Section~\ref{subsec-bao} we will discuss the impact of different parameter choices on geometric observables in light of DESI BAO observations, focusing in particular on the implications of the $\summnu$-$A^{\phi\phi}_\text{lens}$ correlation.

\subsection{The impact of $A^{\phi\phi}_\text{lens}$ and $A^{\mathrm{TTTEEE}}_\text{lens}$ }\label{subsec-alens}
In Figure~\ref{fig:alens_1d_sroll} we show the 1D marginalized posterior distributions for $\summnu$, $n_s$, $A^{\mathrm{TTTEEE}}_\text{lens}$ and $A^{\phi\phi}_\text{lens}$, when both lensing amplitude parameters are allowed to vary. Blue curves correspond to CMB+\texttt{P-ACT} data, while orange curves include BAO data from \texttt{DESI}. Dashed lines represent the fixed-mass case ($\sum m_\nu = 0.06~\mathrm{eV}$), while solid lines indicate configurations where the mass is allowed to vary to negative values.
For CMB+\texttt{P-ACT} with fixed mass (dashed blue) we find a mild deviation from unity ($\sim 1.3\sigma$) in $A^{\phi\phi}_\text{lens}$, while $A^{\mathrm{TTTEEE}}_\text{lens}$ is fully consistent with 1. When $\summnu$ is allowed to vary (solid blue), this trend is reduced, as the additional parameter space broadens the uncertainties absorbing part of the anomaly. As explained in Section~\ref{sec:intro}, in the absence of geometrical information from BAO, $\summnu$ is weakly constrained. When BAO data are added while keeping the neutrino mass fixed (dashed orange), $n_s$ increases, in agreement with~\cite{ACT:2025fju}. In this configuration, the lensing anomaly also becomes more pronounced: $A^{\phi\phi}_\text{lens}$ deviates from unity by $2\sigma$, and $A^{\mathrm{TTTEEE}}_\text{lens}$ by about $1.3\sigma$. Allowing $\summnu$ to vary (solid orange) mitigates this effect, bringing both lensing amplitude parameters within $1\sigma$ from unity. The inferred neutrino mass (best-fit) is $\summnu=-0.018^{+0.085}_{-0.089}(-0.001)~\text{eV}$ and remains compatible with the NO limit at the $0.9\sigma$ C.L.
We conclude that the residual anomaly increases when \texttt{P-ACT} is added to the CMB data and grows further with the addition of \texttt{DESI}. The effect is more prominent in $A^{\phi\phi}_\text{lens}$. This behavior, illustrated by the dashed posterior distributions in Figure~\ref{fig:alens_1d_sroll}, is consistent with the findings of~\cite{Murgia:2020ryi}.

In Figure~\ref{fig:alens_2d_sroll}, we examine the degeneracies among $A_s$, $\summnu$, $n_s$ and $A^{\phi\phi}_\text{lens}$. We compare cases where both lensing amplitude parameters are either fixed to unity (empty contours) or allowed to vary simultaneously (filled contours). Solid contours indicate scenarios with free neutrino mass, while dashed contours correspond to $\sum m_\nu = 0.06~\mathrm{eV}$. For the fixed-mass case, allowing the lensing parameters to vary (blue dashed) shifts the $A_s$ posterior distribution downward compared to keeping them fixed (purple dashed): a higher $A^{\phi\phi}_\text{lens}$ increases the amplitude lensing potential, which must be compensated by a lower amplitude of the primordial power spectrum $A_s$. Overall, introducing the lensing parameters does not significantly weaken the constraints.
From Figure~\ref{fig:alens_2d_sroll} one notices that when the only extra free parameter is $\summnu$ (solid green), there is a positive correlation between $A_s$ and $\summnu$, along with a milder one between $n_s$ and $\summnu$ (see also the green contour of Figure~\ref{fig:high_l} in Appendix~\ref{ap:app-planck}). 
Interestingly, including $A^{\phi\phi}_\text{lens}$ (solid orange) reverses this trend: since both $\summnu$ and $A^{\phi\phi}_\mathrm{lens}$ affect the lensing potential power spectrum in a similar way, allowing $A^{\phi\phi}_\text{lens}$ to vary changes the $A_s$–$\summnu$ degeneracy into a negative correlation.
As a consequence, the dominant degeneracy shifts from $A_s$–$n_s$ to $\summnu$–$A^{\phi\phi}_\text{lens}$. 
This behavior is in line with the findings of \cite{Craig:2024tky, Green:2024xbb}, where only the impact on the scaling of the lensing potential due to negative neutrino masses was considered.
\begin{figure}[b]
    \centering
    \includegraphics[width=0.5\textwidth]{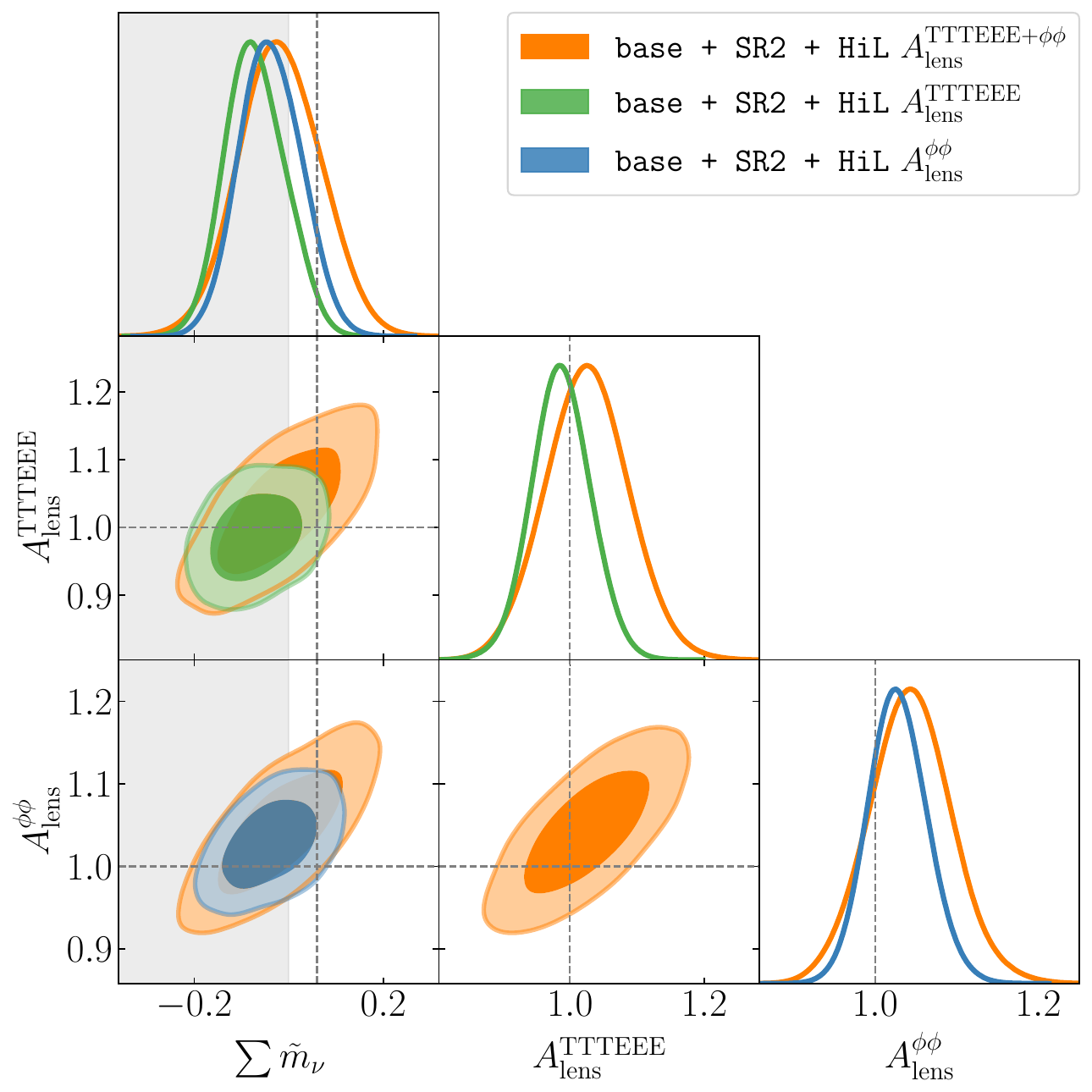}
    \caption{Constraints on $\summnu$, $A^{\phi\phi}_\text{lens}$ and $A^{\mathrm{TTTEEE}}_\text{lens}$. All analyses include the \texttt{base} data sets (DESI DR2 BAO, \texttt{P-ACT} lensing, low-$\ell$ TT from \texttt{Commander}), with the addition of \texttt{SRoll2} and \texttt{HiLLiPoP}. The neutrino mass is allowed to vary. Orange contours correspond to the case where both lensing amplitude parameters are varied, blue contours to the case with $A^{\mathrm{TTTEEE}}_\text{lens}=1$, and green contours to the case with $A^{\phi\phi}_\text{lens}=1$. The grey shaded region indicates the unphysical parameter space.}
    \label{fig:triangle}
\end{figure}

\begin{figure*}[t]
    \centering
    \includegraphics[width=\textwidth]{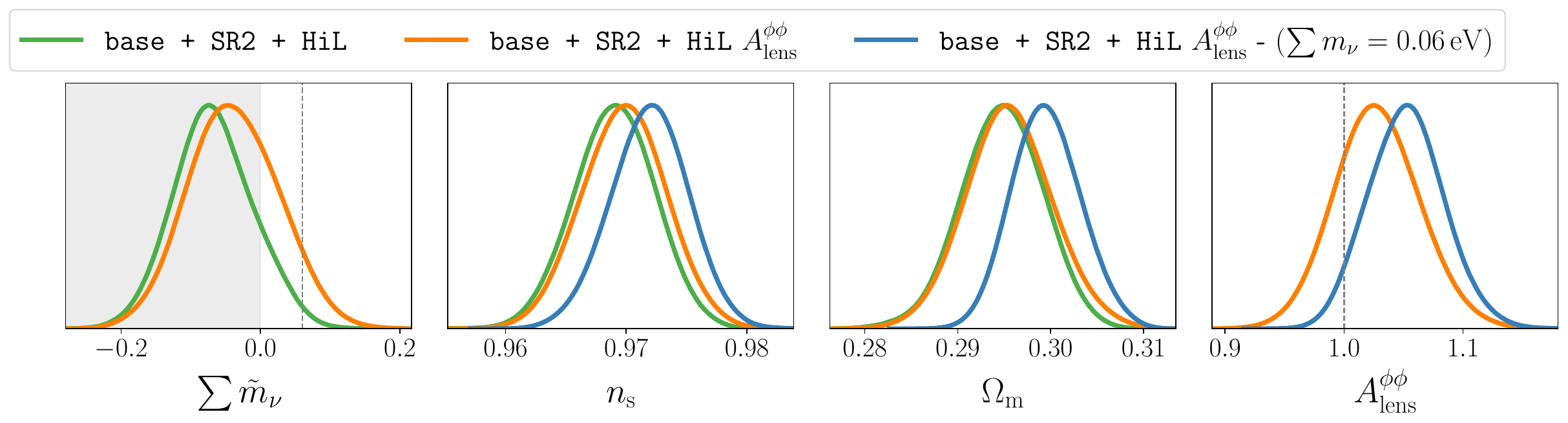}
    \caption{1D marginalized posterior distribution for $\summnu$, $n_s$ $\Omega_\mathrm{m}$ and $A^{\phi\phi}_\text{lens}$. All analyses include the \texttt{base} data sets (DESI DR2 BAO, \texttt{P-ACT} lensing, low-$\ell$ TT from \texttt{Commander}) with the addition of \texttt{SRoll2} and \texttt{HiLLiPoP}. The green line shows the posterior distribution for the case where only the neutrino mass is varied in addition to the \lcdm parameters. The orange line corresponds to the case in which both $\summnu$ and $A^{\phi\phi}_\text{lens}$ are treated as free parameters, while the blue line depicts the case in which $A^{\phi\phi}_\text{lens}$ is varied but the neutrino mass is fixed to the NO limit.
    Vertical grey dashed lines mark the NO mass limit and the fiducial value of unity for the lensing anomaly parameter. The grey shaded region indicates the unphysical parameter space.}
    \label{fig:alens_swap}
\end{figure*}

In Figure~\ref{fig:triangle} we present a triangle plot which complements the information contained in Figure~\ref{fig:alens_2d_sroll}, highlighting the correlation between $\summnu$ and the lensing parameters. Orange contours show the case in which both $A^{\phi\phi}_\text{lens}$ and $A^{\mathrm{TTTEEE}}_\text{lens}$ vary. Green (blue) contours correspond to scenarios where $A^{\phi\phi}_\text{lens}$ ($A^{\mathrm{TTTEEE}}_\text{lens}$) is fixed to unity.
When both lensing amplitude parameters are free, the $\summnu$ posterior distribution is consistent with the NO lower limit at $\lesssim 1\sigma$ significance, and both $A^{\phi\phi}_\text{lens}$ and $A^{\mathrm{TTTEEE}}_\text{lens}$ are fully consistent with unity.
If only $A^{\mathrm{TTTEEE}}_\text{lens}$ varies, there is no shift of the $\summnu$ posterior distribution toward positive values ($\summnu=-0.071^{+0.059}_{-0.068}~\text{eV}$) with respect to the case with fixed lensing parameters (see Table~\ref{tab:part1} in Appendix~\ref{ap:app-additional}). Conversely, allowing only $A^{\phi\phi}_\text{lens}$ to vary shifts the posterior distribution toward positive values ($\summnu=-0.038^{+0.064}_{-0.068}~\text{eV}$). 
We quantify the correlation between parameters A and B using the Pearson correlation coefficient $\mathcal{P}(\text{A–B})\in[-1,1]$, where a higher absolute value indicates stronger correlation. We find $\mathcal{P}(\summnu$–$A^{\phi\phi}_\text{lens}) = 0.47$ and $\mathcal{P}(\summnu$–$A^{\mathrm{TTTEEE}}_\text{lens}) = 0.29$. We conclude that the observed shift in the neutrino mass posterior distribution is primarily driven by the $\summnu$–$A^{\phi\phi}_\text{lens}$ degeneracy.

To better investigate this degeneracy -- and given that $A^{\mathrm{TTTEEE}}_\text{lens}$ is not anomalous -- in Figure~\ref{fig:alens_swap} we explicitly show the trade-off between the neutrino mass anomaly and the extra-power in the lensing potential, by fixing $A^{\mathrm{TTTEEE}}_\text{lens}=1$. The 1D marginalized posterior distributions are shown for $\summnu$, $n_s$, $\Omega_\mathrm{m}$, and $A^{\phi\phi}_\text{lens}$. Green lines correspond to the baseline configuration CMB+\texttt{P-ACT + DESI} with free neutrino mass. Orange lines includes variation of $A^{\phi\phi}_\text{lens}$, while blue lines shows the case with $\sum m_\nu$ fixed at the NO limit and $A^{\phi\phi}_\text{lens}$ free.
In the standard scenario (green), the neutrino mass deviates from the NO limit at $2.2\sigma$. We report $\summnu = -0.068^{+0.054}_{-0.061}~\text{eV}$,  $n_s = 0.9690^{+0.0033}_{-0.0034}$ and $\Omega_\mathrm{m} = 0.2948\pm0.0044$ (see Table~\ref{tab:part1} in Appendix~\ref{ap:app-additional}). 
When both $\summnu$ and $A^{\phi\phi}_\text{lens}$ are free (orange), the posterior distribution shifts toward positive values, reducing the deviation in the neutrino mass to approximately $1.5\sigma$. In this case, $A^{\phi\phi}_\text{lens}$ is consistent with unity. No significant shifts are observed for $n_s$ or $\Omega_\mathrm{m}$. We report $\summnu =  -0.038^{+0.054}_{-0.068}~\text{eV}$,  $n_s = 0.9698^{+0.0035}_{-0.0036}$ and $\Omega_\mathrm{m} = 0.2955\pm0.0047$ (see Table~\ref{tab:part2} in Appendix~\ref{ap:app-additional}).
In the scenario with fixed $\sum m_\nu = 0.06~\mathrm{eV}$, $n_s$ and $\Omega_\mathrm{m}$  slightly increase, to $n_s = 0.9719\pm0.0032$ and $\Omega_\mathrm{m} = 0.2995^{+0.0036}_{-0.0039}$ respectively. 
When the neutrino mass is allowed to vary, including negative values, the inferred $\Omega_\mathrm{m}$ is lower than that obtained from CMB-only analyses. This occurs because the negative neutrino mass absorbs part of the tension between the CMB and BAO measurements. In contrast, when the mass is fixed but the lensing amplitude $A^{\phi\phi}_\text{lens}$ is free to vary, the $\Omega_\mathrm{m}$ prediction shifts towards the CMB-inferred value-- although slightly, from 2.8 to 2.5$\sigma$ C.L. (using the $\Omega_\mathrm{m}$ value reported in \cite{Tristram:2023haj}).
As discussed earlier, when fixing the neutrino mass the tension previously regarding $\summnu$ reappears as an enhancement in $A^{\phi\phi}_\text{lens}$, which now exhibits a $1.7\sigma$ deviation from unity\footnote{Note however that the $A^{\phi\phi}_\text{lens}$--$\summnu$ mapping is not unique, due to the complexity of the parameter interconnection~\cite{Sharma:2025iux}.}.
This trade-off becomes particularly apparent when the mass is fixed to positive values. In this case, there is no way to reconcile CMB and BAO data by adjusting $\summnu$: the tension in the angular scale of the sound horizon at recombination (or equivalently in $\Omega_\mathrm{m}$) cannot be alleviated. As a result, the discrepancy between early- and late-time measurements of cosmological observables manifests as an apparent excess in the amplitude of the lensing potential.

\begin{figure*}[t]
    \centering
    \includegraphics[width=\textwidth]{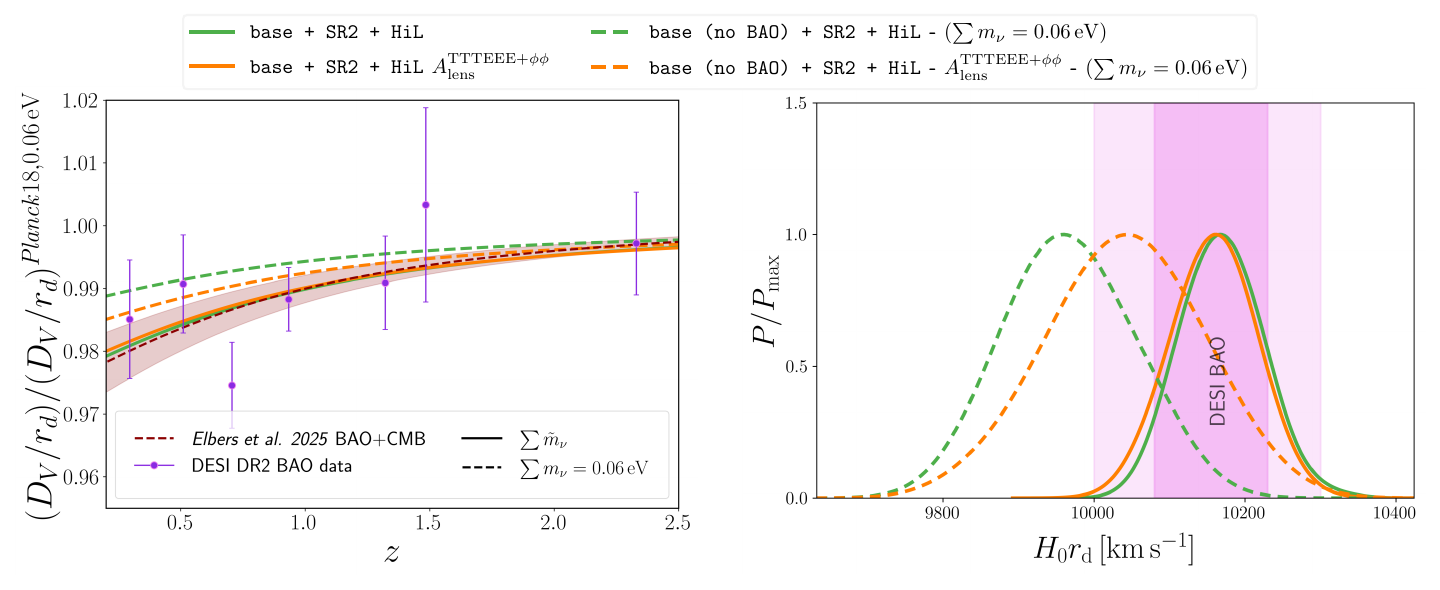}
    \caption{\textit{Left:} Best-fit reconstruction of the isotropic BAO distance, normalized by its value in $\Lambda$CDM. The purple points correspond to the DESI DR2 BAO measurements reported in Table IV of \cite{DESI:2025zgx}. All analyses include \texttt{P-ACT lensing}, low-$\ell$ TT data from \texttt{Commander}, low-$\ell$ EE data from \texttt{SRoll2}, and high-$\ell$ TTTEEE data from \texttt{HiLLiPoP}. Solid lines include BAO data from DESI DR2, while dashed lines exclude this data set and fix the neutrino mass to the NO lower limit. Orange curves correspond to analyses with varying lensing amplitude parameters; green curves represent cases where these parameters are fixed. The brown dashed line and shaded region represents the mean and $68\%$ C.L. of the baseline BAO+CMB result from \cite{DESI:2025ejh}, where the CMB data set corresponds to the combination \texttt{base + SimAll + CamSpec}. This configuration yields a value of $\sum m_\nu = -0.101~\mathrm{eV}$. \textit{Right:} 1D posterior distribution of $H_\mathrm{0}r_d$ for the same configurations shown in the left panel.  The vertical shaded bands indicate the  68$\%$ and 95$\%$ confidence regions from the DESI DR2 BAO data.}
    \label{fig:Dv_H0_rd}
\end{figure*}

\subsection{Implications for BAO in light of DESI}\label{subsec-bao}

In the left panel of Figure~\ref{fig:Dv_H0_rd} we show our best-fit curve for the isotropic BAO distance, \textit{i.e.}, the spherically-averaged volume distance $D_V(z)$\footnote{$D_V(z)=\left(z D_M(z)^2 D_H(z)\right)^{1/3}$ with $D_M(z)$ being the transverse comoving distance and $D_H(z)$ the line-of-sight distance~\cite{DESI:2025zgx}.}, divided by the sound horizon at the baryon drag epoch, $r_d(z)$.
We present results for the \texttt{SRoll2 + HiLLiPoP} configuration with (solid orange) and without (solid green) varying the lensing amplitude parameters. 
We also include cases where the neutrino mass is fixed and DESI DR2 BAO data are excluded from the analysis (dashed lines). In the cases presented, both $A^{\mathrm{TTTEEE}}_\text{lens}$ and $A^{\phi\phi}_\text{lens}$ are free to vary. This set-up is chosen because it produces the neutrino mass posterior distribution most consistent with $\sum m_\nu = 0.06~\mathrm{eV}$ (see Section~\ref{subsec-alens}).
For comparison, the brown dashed line and shaded region shows the mean and $1\sigma$ C.L. from \cite{DESI:2025ejh} (see their Figure 12\footnote{They were obtained by fitting DESI DR2 BAO together with \texttt{P-ACT} lensing data and their baseline CMB likelihoods (\texttt{Commander + SimAll + CamSpec}) under a model with a varying effective neutrino mass.}).
All curves are normalized to the \lcdm \textit{Planck18}~\cite{Planck:2018vyg} best-fit prediction with $\sum m_\nu = 0.06~\mathrm{eV}$.
The configuration with free lensing parameters and fixed mass (dashed orange) does not include the BAO likelihood. Nevertheless, its best-fit expansion history, expressed in $D_V/r_d$, is in good qualitative agreement with the results obtained by~\cite{DESI:2025ejh}, which in their case arise under the assumption of a strongly negative neutrino mass (with lensing parameters fixed to unity). 
In contrast, the case without lensing parameter variation (dashed green) shows a larger discrepancy with the BAO data.
We can conclude that varying $A^{\phi\phi}_\text{lens}$ helps reconcile CMB predictions with BAO observations.

These results reinforce the interpretation that the apparent preference for negative neutrino masses is pointing to deeper internal inconsistencies across early-time (\textit{Planck} CMB TTTEEE) and late-time (\texttt{P-ACT} lensing and DESI BAO) probes. With a careful choice of CMB likelihoods, such as \texttt{HiLLiPoP} and \texttt{SRoll2}, it is possible to provide a good fit of both CMB and BAO data without invoking unphysical values of $\sum m_\nu$, though this comes at the cost of accepting a residual -- still unexplained -- enhanced amplitude of the lensing potential power spectrum.
The right panel of Figure~\ref{fig:Dv_H0_rd} further illustrates this point, showing the 1D marginalized posterior distributions of the parameter combination $H_\mathrm{0} r_d$ for the same configurations shown in the left panel. The \texttt{SRoll2 + HiLLiPoP} with and without free lensing amplitude parameters (green and orange solid lines), show excellent agreement with the BAO measurements, as was also the case for $D_V/r_d$. They fall entirely within the BAO $1\sigma$ C.L., failing however to be fully consistent with the NO mass limit (see Table~\ref{tab:part1} and \ref{tab:part2} in Appendix~\ref{ap:app-additional}). 
As expected, when BAO data are not included in the analysis, the tension with DESI is maximum in the \lcdm $+ \sum m_\nu = 0.06$ eV configuration, as qualitatively indicated by the dashed green line.
On the contrary, the configuration \texttt{SRoll2 + HiLLiPoP + P-ACT} with free lensing parameters (dashed orange) is compatible within $1\sigma$ with the DESI measurement of $H_\mathrm{0} r_d$, due to the capability of $A^{\phi\phi}_\text{lens}$ to partially compensate for the impact of having imposed a positive neutrino mass.

\begin{figure*}[t]
    \centering
    \includegraphics[width=\textwidth]{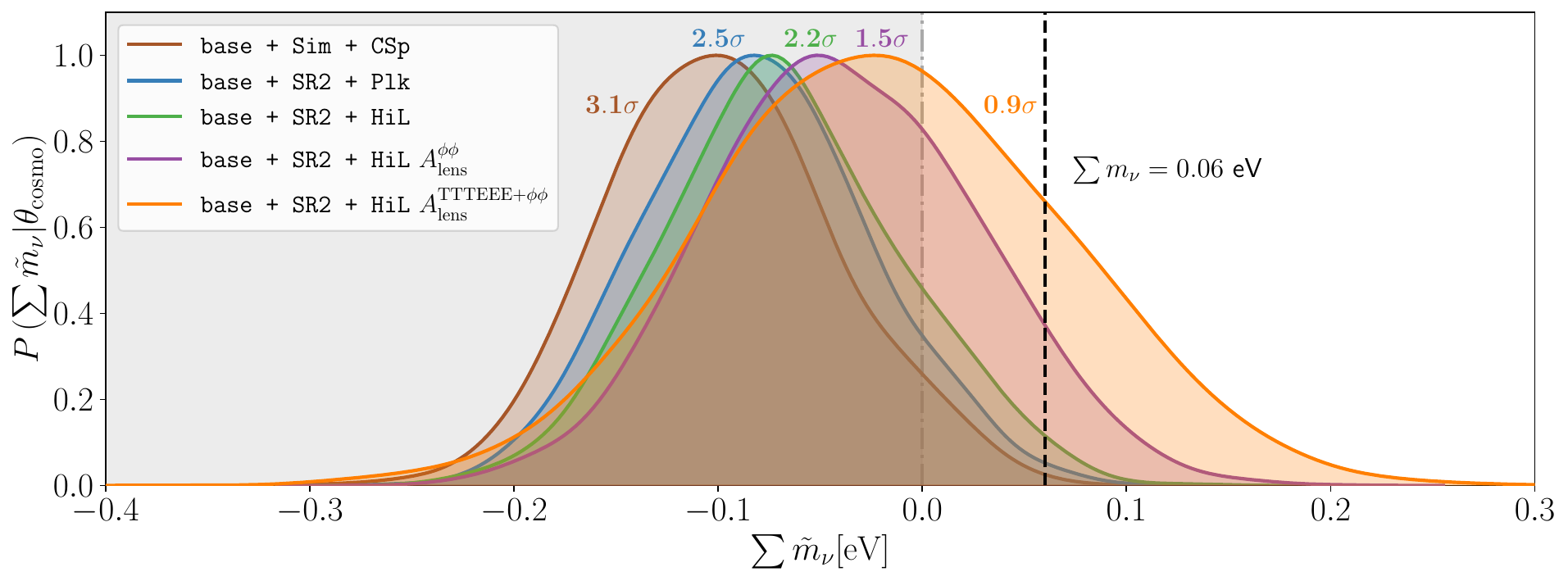}
    \caption{1D marginalized posterior distribution for $\summnu$ under different likelihood configurations. All analyses include DESI DR2 BAO, \texttt{P-ACT lensing}, low-$\ell$ TT data from \texttt{Commander}, color-coded as follows: brown for \texttt{SimAll + CamSpec}
    blue for \texttt{SRoll2 + Plik}, green for \texttt{SRoll2 + HiLLiPoP}, purple for \texttt{SRoll2 + HiLLiPoP} with varying $A^{\phi\phi}_\text{lens}$ and orange \texttt{SRoll2 + HiLLiPoP} with varying $A^{\phi\phi}_\text{lens}$ and $A^{\mathrm{TTTEEE}}_\text{lens}$. The dashed black line indicates the reference value $\sum m_\nu = 0.06~\mathrm{eV}$, with each distribution annotated by its corresponding $\sigma$-level deviation from this value. The grey shaded area marks the unphysical region of the parameter space.}
    \label{fig:mnu_distrib_sigma}
\end{figure*}

\begin{table*}[ht]

\caption{Consistency levels of the neutrino mass 1D marginalized posterior distributions for various configurations,
evaluated with respect to three reference values: $\sum m_\nu = 0$, $0.06$, and $0.10~\mathrm{eV}$. All the reported data include our \texttt{baseline} configuration, namely DESI DR2 BAO, \texttt{P-ACT lensing}, low-$\ell$ TT data from \texttt{Commander} and \texttt{SRoll2 + HiLLiPoP}. From left to right, the first column reports the case of no variation of the lensing amplitude parameters, the second column reports the case in which $A^{\phi\phi}_\mathrm{lens}$ is varied and $A^{\mathrm{TTTEEE}}_\mathrm{lens}=1$, while the last one is when both are allowed to vary freely. For each case, results are shown both for the analysis allowing $\summnu$ to take negative values and for the physically motivated scenario where the prior $\sum m_\nu > 0$ is imposed. We report the consistency levels in terms of Gaussian-equivalent $\sigma$ values, with the corresponding CDF values shown in parentheses, obtained as explained in Section~\ref{sec:data}.}
\centering
\begin{tabular}{l@{\hskip 12pt}l@{\hskip 25pt}c@{\hskip 30pt}c@{\hskip 30pt}c@{\hskip 10pt}}
\hline
\noalign{\vskip 0.5ex}
& \textbf{Ref. value} & \texttt{baseline} & \texttt{baseline} - $A^{\phi\phi}_\mathrm{lens}$ & \texttt{baseline} - $A^{\mathrm{TTTEEE}+\phi\phi}_\mathrm{lens}$ \\
\noalign{\vskip 0.3ex}
\hline
\noalign{\vskip 0.8ex}
\multirow{3}{*}{$\sum \tilde{m}_\nu$}
 & $\sum m_\nu = 0~\mathrm{eV}$ & 1.1$\sigma$ (86.9\%) & 0.6$\sigma$ (72.0\%) & 0.2$\sigma$ (59.2\%) \\

\noalign{\vskip 0.8ex}
 & $\sum m_\nu = 0.06~\mathrm{eV}$ & 2.2$\sigma$ (98.6\%) & 1.5$\sigma$ (92.9\%) & 0.9$\sigma$ (81.1\%) \\

\noalign{\vskip 0.8ex}
 & $\sum m_\nu = 0.1~\mathrm{eV}$ & 3.0$\sigma$ (99.8\%) & 2.1$\sigma$ (98.2\%) & 1.3$\sigma$ (90.9\%) \\
\noalign{\vskip 0.8ex}
\hline
\noalign{\vskip 0.8ex}
\hline
\noalign{\vskip 0.8ex}
\multirow{2}{*}{$\sum m_\nu > 0$}
 & $\sum m_\nu = 0.06~\mathrm{eV}$ & 1.2$\sigma$ (88.4\%) & 0.7$\sigma$ (74.5\%) & 0.1$\sigma$ (54.2\%) \\

\noalign{\vskip 0.8ex}
 & $\sum m_\nu = 0.1~\mathrm{eV}$ & 2.3$\sigma$ (98.9\%) & 1.6$\sigma$ (94.1\%) & 0.8$\sigma$ (78.5\%) \\
\hline
\noalign{\vskip 0.8ex}

\label{tab:summary}
\end{tabular}
\end{table*}

\section{Summary and Conclusion}\label{sec:conclusion}

Our main results are summarized in Figure~\ref{fig:mnu_distrib_sigma} and Table~\ref{tab:summary}.
In Figure~\ref{fig:mnu_distrib_sigma} we show the 1D marginalized posterior distributions for $\summnu$, along with its statistical consistency with the NO lower bound, \textit{i.e.} $\sum m_\nu = 0.06~\mathrm{eV}$, computed using the method described in Section~\ref{sec:data}. Results are shown for the combinations \texttt{base + SimAll + CamSpec} and \texttt{base + SimAll + Plik}, as well as for \texttt{base + SRoll2 + HiLLiPoP}, for which we consider three cases: both lensing amplitude parameters fixed, only $A^{\mathrm{TTTEEE}}_\mathrm{lens}$ fixed, and both parameters free to vary. Table~\ref{tab:summary} presents the significance levels for the last three of these configurations, evaluated against three reference values of the neutrino mass: $\sum m_\nu = 0$, $0.06$, and $0.10~\mathrm{eV}$, which correspond to the minimal values allowed by neutrino oscillation experiments assuming normal and inverted mass ordering, respectively.
For each configuration, we report results with and without the physical prior $\sum m_\nu > 0$. The significance is quoted as Gaussian-equivalent $\sigma$ values, with corresponding CDFs shown in parentheses (see Section~\ref{sec:data}).

This work, like many analyses following the DESI DR2 data release, highlights the deep interconnections between multiple cosmological tensions. We dub this as a ``short blanket” effect: addressing one apparent problem often uncovers another, so that it is very difficult to fully resolve all discrepancies with minimal modifications of the \lcdm model.
We have focused on the apparent preference for negative neutrino masses emerging from different combinations of CMB and BAO data. By systematically exploring combinations of \textit{Planck} high- and low-$\ell$ likelihoods, DESI DR2 BAO, and \textit{Planck}-ACT lensing measurements, we have found this preference to be a persistent feature.
We have analyzed the two distinct impacts of gravitational lensing on the CMB, namely $(i)$ rescaling of lensing potential power spectrum inferred from the CMB trispectrum, and $(ii)$ smoothing of the high-$\ell$ TTTEEE power spectra.
We have described the two effects by introducing the two extra-parameters: $(i)$ $A^{\phi\phi}_\mathrm{lens}$ and $(ii)$ $A^{\mathrm{TTTEEE}}_\mathrm{lens}$.
We have shown that a residual lensing anomaly arises when \texttt{P-ACT} CMB lensing and DESI BAO data are added to an otherwise anomaly-free CMB likelihood. This reflects an inconsistency between early- and late-time cosmological observables, with the lensing amplitude parameter $A^{\phi\phi}_\mathrm{lens}$ deviating up to $2\sigma$ as more late-time data are included. 

We have found that the anomaly primarily resides in $A^{\phi\phi}_\mathrm{lens}$, while $A^{\mathrm{TTTEEE}}_\mathrm{lens}$ remains consistent with unity.
As discussed in detail in~\cite{Sharma:2025iux}, estimates of the neutrino mass are influenced both by measurements of the angular scale of the sound horizon at recombination $\theta_s$, and by the lensing impact on the CMB power spectrum and trispectrum. Focusing on the correlations among the three parameters governing these effects ($\summnu$, $A^{\mathrm{TTTEEE}}_\mathrm{lens}$ and $A^{\phi\phi}_\mathrm{lens}$), we have shown that the apparent preference for a negative neutrino mass -- mostly driven by a discrepancy in the measurement of the angular size of the sound horizon between the CMB and BAO epochs -- can be recast to an anomaly in $A^{\phi\phi}_\mathrm{lens}$. The latter parameter effectively traces the extra lensing power required to provide a good joint fit to \texttt{P-ACT} lensing, DESI BAO data, and $Planck$ CMB power spectra.

The tension between $A^{\phi\phi}_\mathrm{lens}$ (anomalous) and $A^{\mathrm{TTTEEE}}_\mathrm{lens}$ (compatible with 1) seems to point to a discrepancy between early- (\textit{Planck} CMB TTTEEE) and late-time (\texttt{P-ACT} lensing and DESI BAO) cosmological measurements. 
Looking forward, combining multiple probes to break parameter degeneracies will be crucial to shed light on the source of this discrepancy. In addition, upcoming large-scale structure data will be the key to obtaining tight and reliable estimates of the absolute neutrino mass scale~\cite{Euclid:2024imf}.

\begin{acknowledgements}

We are very thankful to Stefano Gariazzo, Carlo Giunti, Julien Lesgourgues, Luca Pagano and Vivian Poulin for useful discussions.
A.C. thanks the INFN Section, the University, and the INAF Astronomical Observatory of Cagliari (Sardinia) for their hospitality during the
development of this project. We also thank the Director and the Computing and Network Service of Laboratori Nazionali del Gran Sasso (LNGS-INFN).
We acknowledge the use of {\tt Matplotlib}~\cite{Hunter:2007ouj}, {\tt NumPy}~\cite{Harris:2020xlr}, {\tt SciPy}~\cite{Virtanen:2019joe}, {\tt GetDist}~\cite{Lewis:2019xzd}.

\end{acknowledgements}

\bibliography{bibliography}

\appendix

\section{The CMB lensing amplitude parameters}\label{ap:app-Alens}

In this Appendix, we further clarify the use of the parameters $A_\text{lens}$, $A^{\mathrm{TTTEEE}}_\text{lens}$ and $A^{\phi\phi}_\text{lens}$ in our version of \texttt{CLASS}~\cite{Simard:2017xtw,Murgia:2020ryi,Corona:2021qxl}, and the associated theoretical framework~\cite{Simard:2017xtw}. These parameters appear in the \texttt{harmonic} and \texttt{lensing} modules of \texttt{CLASS}~\cite{Lesgourgues2011CosmicI, Blas2011Cosmic}. 

The parameter $A_\text{lens}$~\cite{Calabrese:2008rt,Simard:2017xtw} affects both the reconstruction of the lensing potential power spectrum and the smearing of the acoustic peaks. The parameter $A^{\phi\phi}_\text{lens}$, scales only the amplitude of the lensing potential power spectrum, reconstructed from the CMB four-point correlation functions. The CMB lensing anomaly is effectively described by the fact that both these parameters deviates from unity, when let free to vary to fit the full set of \textit{Planck} data.
The lensed temperature and polarization angular power spectra can be written as convolutions of the unlensed fields $C^X_\ell$ with the lensing potential $C^{\phi\phi}_\ell$ ($X \in \{\text{TT, TE, EE}\}$). This allows us to define a third parameter, dependent on the two previously introduced\footnote{Note that, in all our analyses, when two lensing amplitude parameters are free to vary the third one is set to unity}, as

\begin{equation}\label{eq:lensingdef}
A^{\mathrm{TTTEEE}}_\text{lens} = A_\text{lens} \times A^{\phi\phi}_\text{lens}\, ,
\end{equation}

so that the smearing effect of the CMB peaks is fully described by $A^{\mathrm{TTTEEE}}_\text{lens}$.

We now give a short and pedagogical review of the formalism behind the introduction of the three lensing amplitude parameters defined in Equation~\ref{eq:lensingdef}. For a more detailed and comprehensive discussion we refer to~\cite{Lewis:2006fu}.

Working at the lowest expansion order in $C^{\phi\phi}_\ell$ (see Section 4.1 in \cite{Lewis:2006fu}), we focus on the temperature anisotropy field (analogous considerations apply to E- and B-modes). We assume a statistically isotropic unlensed temperature field $\Theta(\textbf{x})\equiv \delta T($\textbf{x}$)/\bar{T}$, where $\textbf{x}$ denotes the direction on the sky at which the temperature anisotropy is measured and $\bar{T}$ is the mean temperature. In Fourier space, the temperature anisotropy field can be represented by its Fourier transform $\Theta(\boldsymbol{\ell})$. The two-point correlation function between directions $\mathbf{x}$ and $\mathbf{x}'$ is then
\begin{equation}
    C^{\Theta}_\ell \delta(\boldsymbol{\ell}-\boldsymbol{\ell}')= \langle\Theta(\boldsymbol{\ell})\Theta^*(\boldsymbol{\ell}')\rangle
\end{equation}
where $\boldsymbol{\ell}$ is the multipole vector. Its magnitude, $\ell \equiv |\boldsymbol{\ell}|$, is the multipole number, which corresponds to angular scales on the sky. Due to statistical isotropy, the correlation depends only on the separation $r = |\mathbf{x} - \mathbf{x}'|$ between the two directions.
In a similar fashion, the lensed temperature field can be expressed as $\tilde{\Theta}(\textbf{x}) = \Theta (\textbf{x}+\nabla \phi)$, where at lowest order the CMB photon deflection angle is given by the gradient of the lensing potential $\phi$. In Fourier space, the corresponding two-point correlation function is $\langle\tilde\Theta(\boldsymbol{\ell})\tilde\Theta^*(\boldsymbol{\ell}')\rangle= \tilde{C}^{\Theta}_\ell \delta(\boldsymbol{\ell}-\boldsymbol{\ell}')$. Using a Taylor expansion and the properties $\langle\phi(\boldsymbol{\ell})\phi^*(\boldsymbol{\ell}')\rangle = C^{\phi\phi}_\ell \delta(\boldsymbol{\ell}-\boldsymbol{\ell}')$ and $\phi(\boldsymbol{\ell}) = \phi^*(-\boldsymbol{\ell})$, the lensed angular power spectrum can be written as a sum of two terms: 

\begin{equation}
\begin{aligned}
    \tilde{C}^{\Theta}_\ell \approx C^{\Theta}_\ell + \int \frac{d^2\boldsymbol{\ell}'}{(2\pi)^2} & W_\mathrm{smooth}(\boldsymbol{\ell},\boldsymbol{\ell}')\, C_{|\boldsymbol{\ell}- \boldsymbol{\ell}'|}^{\phi\phi} C^{\Theta}_{\ell'} \\&
    - C^{\Theta}_\ell W_\mathrm{dump}(\ell) \, .
\end{aligned}
\end{equation}

The first one is a smoothing term, which re-destributes power from large to small scales, while the second one is a damping contribution, which lowers the peak contrast\footnote{See Eq.~4.11 of \cite{Lewis:2006fu} for the explicit expression of the $W_\mathrm{smooth}$ and $W_\mathrm{dump}$ terms}.
By introducing the lensing potential amplitude parameter $A^{\phi\phi}_\text{lens}$ in the definition of the lensing potential ($\hat{C}^{\phi\phi}_\ell = A^{\phi\phi}_\text{lens}{C}^{\phi\phi}_\ell$) we can write the variation in the CMB peak smoothing as

\begin{equation}
\begin{aligned}
    &\Delta\tilde{C}^{\Theta}_\ell 
    \approx  A_{\text{lens}}\int \frac{d^2\boldsymbol{\ell}'}{(2\pi)^2}  
    W_\mathrm{smooth}(\boldsymbol{\ell},\boldsymbol{\ell}')\,
    \hat{C}_{|\boldsymbol{\ell}- \boldsymbol{\ell}'|}^{\phi\phi} 
    C^{\Theta}_{\ell'} \\
    &= A_{\text{lens}}\times A^{\phi\phi}_\text{lens}
    \int \frac{d^2\boldsymbol{\ell}'}{(2\pi)^2}  
    W_\mathrm{smooth}(\boldsymbol{\ell},\boldsymbol{\ell}')\,
    C_{|\boldsymbol{\ell}- \boldsymbol{\ell}'|}^{\phi\phi} 
    C^{\Theta}_{\ell'} \\
    &\overset{\text{Eq.}\eqref{eq:lensingdef}}{=} 
    A^{\mathrm{TTTEEE}}_\text{lens}
    \int \frac{d^2\boldsymbol{\ell}'}{(2\pi)^2}  
    W_\mathrm{smooth}(\boldsymbol{\ell},\boldsymbol{\ell}')\,
    C_{|\boldsymbol{\ell}- \boldsymbol{\ell}'|}^{\phi\phi} 
    C^{\Theta}_{\ell'},
\end{aligned}
\end{equation}

where the parameter $A^{\mathrm{TTTEEE}}_\text{lens}$ appears explicitly.

Concerning the implementation in \texttt{CLASS}, the steps are somewhat more involved, yet they yield the same outcome. Practically, the first modification is implemented by applying the multiplicative factor $A_\text{lens} \times A^{\phi\phi}_\text{lens}$ to the lensing potential power spectrum
\begin{equation}
    C^{\phi\phi}_\ell \propto  A_\text{lens} \times A^{\phi\phi}_\text{lens}\int_0^\infty \frac{dk}{k} \, P(k) \left| \Delta_\ell(k) \right|^2\,.
\end{equation}
Then, the factor $A^{\mathrm{TTTEEE}}_\text{lens}/ A^{\phi\phi}_\text{lens}$ is applied to the quantities $C_\mathrm{gl}(r)$ and $C_\mathrm{gl,2}(r)$ computed in \texttt{lensing} modulus. For further details, see Section 4.2 of \cite{Lewis:2006fu}.

\section{The CMB lensing anomaly}\label{ap:app-lensing}
\begin{figure*}[t]
    \centering
    \includegraphics[width=0.85\textwidth]{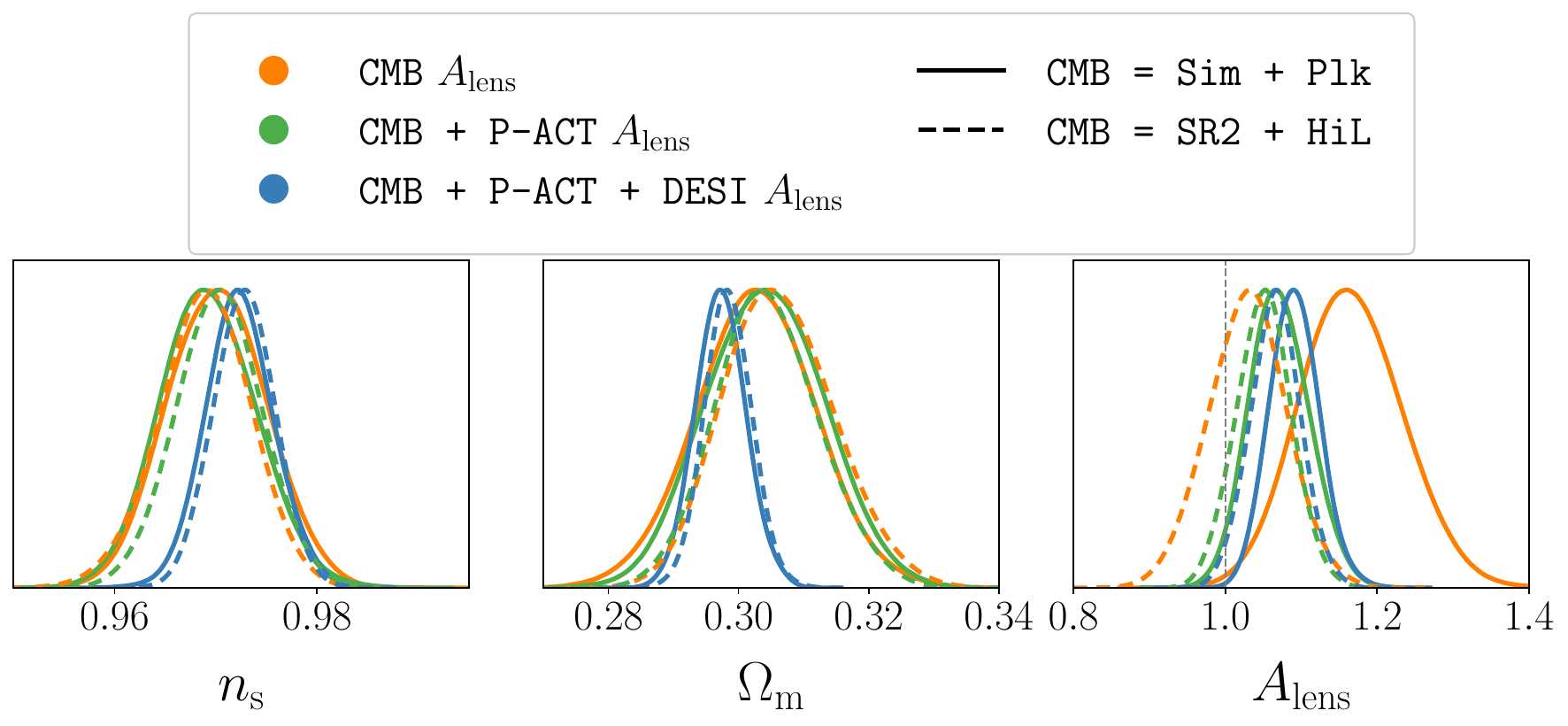}
    \caption{1D marginalized posterior distributions for $n_s$, $\Omega_\mathrm{m}$, and $A_\mathrm{lens}$. Orange lines correspond to the CMB-only case, green lines include \texttt{P-ACT} lensing data, and blue lines additionally include DESI DR2 BAO. Solid lines denote the CMB combination \texttt{SimAll + Plik}, while dashed lines correspond to \texttt{SRoll2 + HiLLiPoP}. Vertical grey dashed lines mark the fiducial value of unity for the lensing anomaly parameter.}
    \label{fig:tot_marginal}
\end{figure*}
\begin{figure*}[t]
    \centering
    \includegraphics[width=\textwidth]{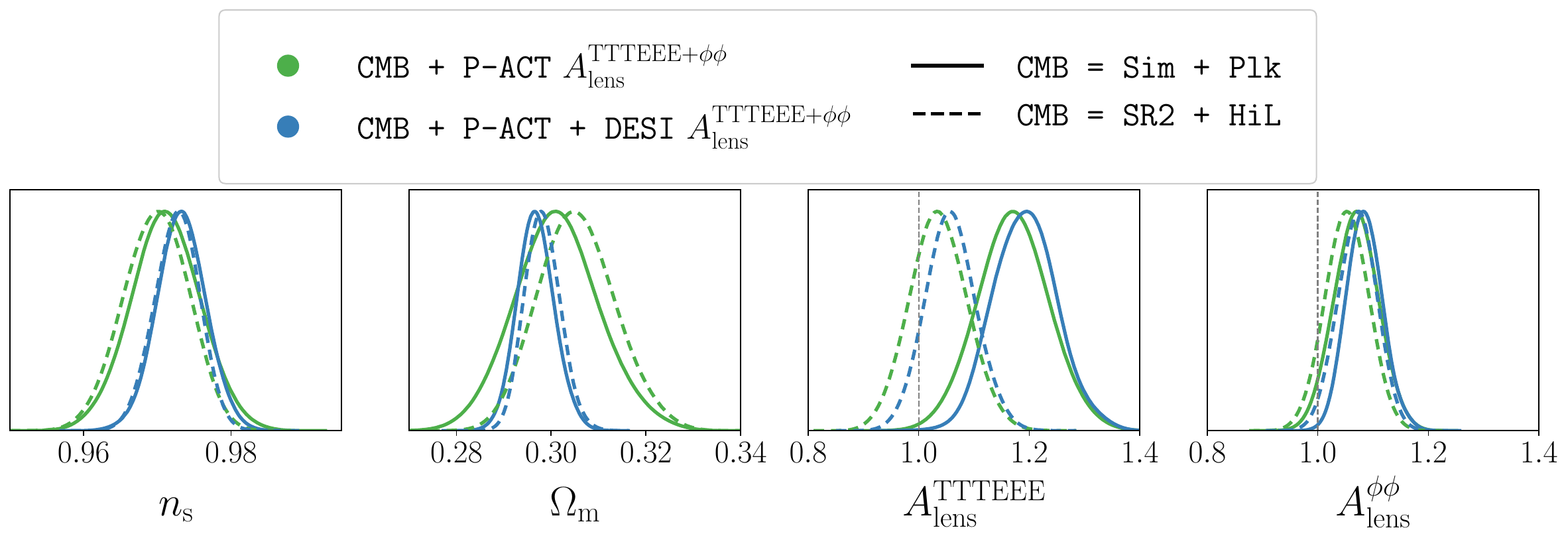}
    \caption{1D marginalized posterior distributions for $n_s$, $\Omega_\mathrm{m}$, $A^{\mathrm{TTTEEE}}_\text{lens}$ and $A^{\phi\phi}_\mathrm{lens}$. Green lines include \texttt{P-ACT} lensing data, and blue lines additionally include DESI DR2 BAO. Solid lines denote the CMB combination \texttt{SimAll + Plik}, while dashed lines correspond to \texttt{SRoll2 + HiLLiPoP}. Vertical grey dashed lines mark the fiducial value of unity for the lensing amplitude parameters.}
    \label{fig:ppTE_marginal}
\end{figure*}
In this Appendix, we review the lensing anomaly and how it appears under different combinations of CMB likelihoods, with and without including lensing potential measurements (\textit{i.e.}, \texttt{P-ACT}) and DESI BAO data. We employ the following CMB likelihoods: the standard PR3 combination of \texttt{Commander} at low-$\ell$ TT, \texttt{SimAll} at low-$\ell$ EE and \texttt{Plik} at high-$\ell$ TTTEEE; as well as the combination of \texttt{Commander} at low-$\ell$ TT, \texttt{SRoll2} at low-$\ell$ EE and \texttt{HiLLiPoP} at high-$\ell$ TTTEEE.

The \texttt{Plik}-legacy likelihood exhibits the largest discrepancy in the lensing amplitude parameter $A_\mathrm{lens}$, deviating by about $2.8\sigma$ from unity~\cite{Calabrese:2008rt, Planck:2018vyg, Addison:2023fqc}. In contrast, the \texttt{HiLLiPoP} reanalysis found a non-anomalous value, only $0.7\sigma$ from unity~\cite{Addison:2023fqc, Tristram:2023haj}. The inferred value of $A_\mathrm{lens}$ changes depending on the data set used in the analysis, as illustrated in Figure~\ref{fig:tot_marginal}. The 1D marginalized posterior distributions are shown for $n_s$, $\Omega_\mathrm{m}$, and $A_\text{lens}$. Solid lines represent results from the \texttt{SimAll + Plik} combination, while dashed lines show those from \texttt{SRoll2 + HiLLiPoP}. Orange lines correspond to CMB-only data, green lines include \texttt{P-ACT} lensing data, and blue lines additionally incorporate DESI BAO data.
The resulting deviations of $A_\mathrm{lens}$ from unity are as follows. For \texttt{SimAll + Plik}, we find $2.7\sigma$, which decreases to $1.9\sigma$ when including \texttt{P-ACT} and rises to $3.1\sigma$ with the addition of DESI. For \texttt{SRoll2 + HiLLiPoP}, the discrepancy is $0.6\sigma$ for CMB-only data, rising to $1.4\sigma$ with the addition of \texttt{P-ACT}, and reaching $2.0\sigma$ when DESI DR2 BAO are also included. We do not observe any significant shifts in $n_s$ or $\Omega_\mathrm{m}$ when using different CMB likelihoods. Including DESI BAO, however, slightly shifts $n_s$ upwards, and improves the constraining power on $\Omega_\mathrm{m}$.

For CMB-only analyses, displaying results with both lensing amplitude parameters is not particularly informative, since $A^{\phi\phi}_\text{lens}$ cannot be constrained without external lensing data, and $A^{\mathrm{TTTEEE}}_\text{lens}$ fully mimics the behavior of $A_\mathrm{lens}$.
As such, in Figure~\ref{fig:ppTE_marginal} 
we present results for the CMB likelihood configurations considered in this work in combination with \texttt{P-ACT} (green) and \texttt{P-ACT + DESI} (blue). The C.L. for $A^{\phi\phi}_\mathrm{lens}$ ($A^{\mathrm{TTTEEE}}_\text{lens}$) consistency with unity are as follows.
For \texttt{SimAll + Plik + P-ACT}, we find $1.7\sigma$ ($2.7\sigma$), increasing to $2.7\sigma$ ($3.2\sigma$) with DESI. For \texttt{SRoll2 + HiLLiPoP} we find $1.3\sigma$ ($0.7\sigma$) with \texttt{P-ACT} and $2.0\sigma$ ($1.3\sigma$) when DESI is added. Thus, the anomaly in $A^{\mathrm{TTTEEE}}_\text{lens}$ is significantly reduced when \texttt{SRoll2 + HiLLiPoP} is used.
Considerations done for $n_s$ and $\Omega_\mathrm{m}$ in Figure~\ref{fig:tot_marginal} apply to Figure~\ref{fig:ppTE_marginal} too.
As already noted in the main text, $A^{\mathrm{TTTEEE}}_\text{lens}$, which captures the phenomenology of $A_\text{lens}$, remains the most anomalous parameter in the \texttt{Plik} dat set. An anomaly also persists in $A^{\phi\phi}_\text{lens}$ for \texttt{Plik}. The use of \texttt{HiLLiPoP} largely mitigates the $A^{\mathrm{TTTEEE}}_\text{lens}$ anomaly. However, a residual anomaly in $A^{\phi\phi}_\text{lens}$ remains.
This residual anomaly increases when \texttt{P-ACT} and DESI data are added to the CMB data sets.

\begin{figure*}[t]
    \centering
    \includegraphics[width=\textwidth]{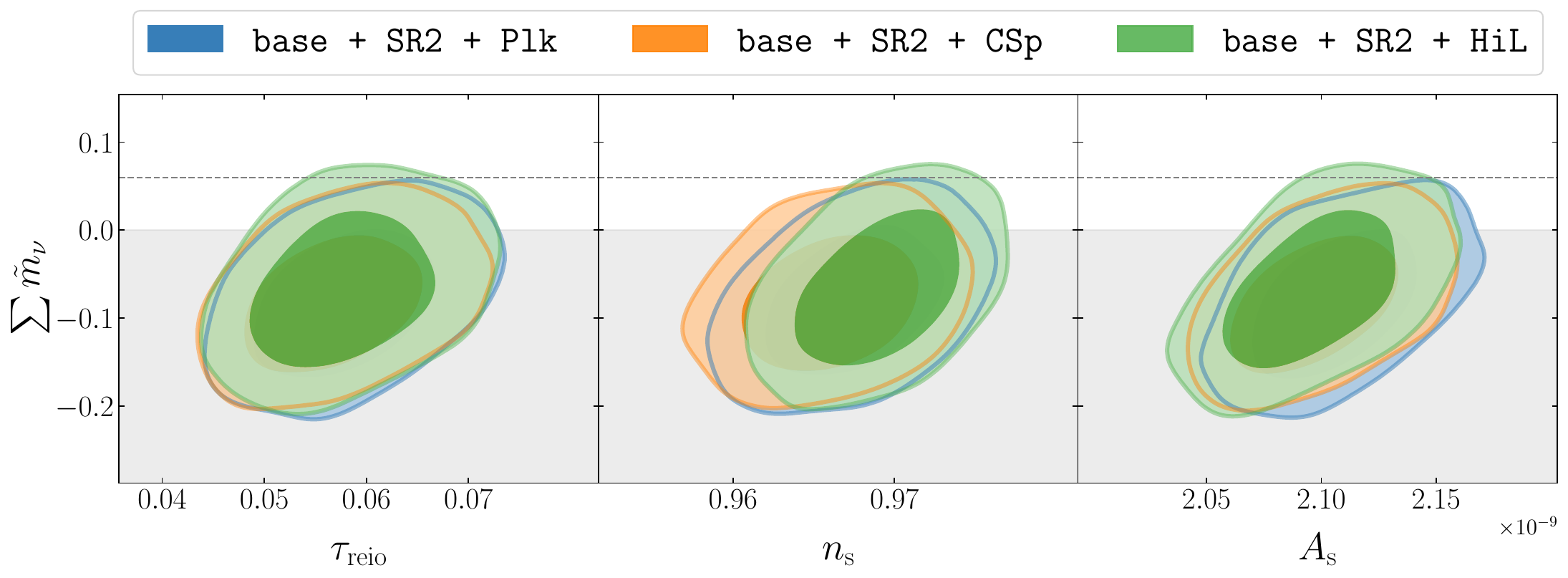}
    \caption{Constraints for $\summnu$, $\tau_\mathrm{reio}$, $n_s$, and $A_s$. All analyses include DESI DR2 BAO, \texttt{P-ACT} lensing, low-$\ell$ TT from \texttt{Commander} (\texttt{base}), and low-$\ell$ EE from \texttt{SRoll2}. Results are shown for \texttt{Plik} (blue), \texttt{CamSpec} (orange), and \texttt{HiLLiPoP} (green). The dashed horizontal line indicates the NO limit for the neutrino mass, $\sum m_\nu = 0.06~\mathrm{eV}$. The grey shaded region denotes the unphysical part of the parameter space.}
    \label{fig:high_l}
\end{figure*}

\begin{figure*}[t]
    \centering
    \includegraphics[width=\textwidth]{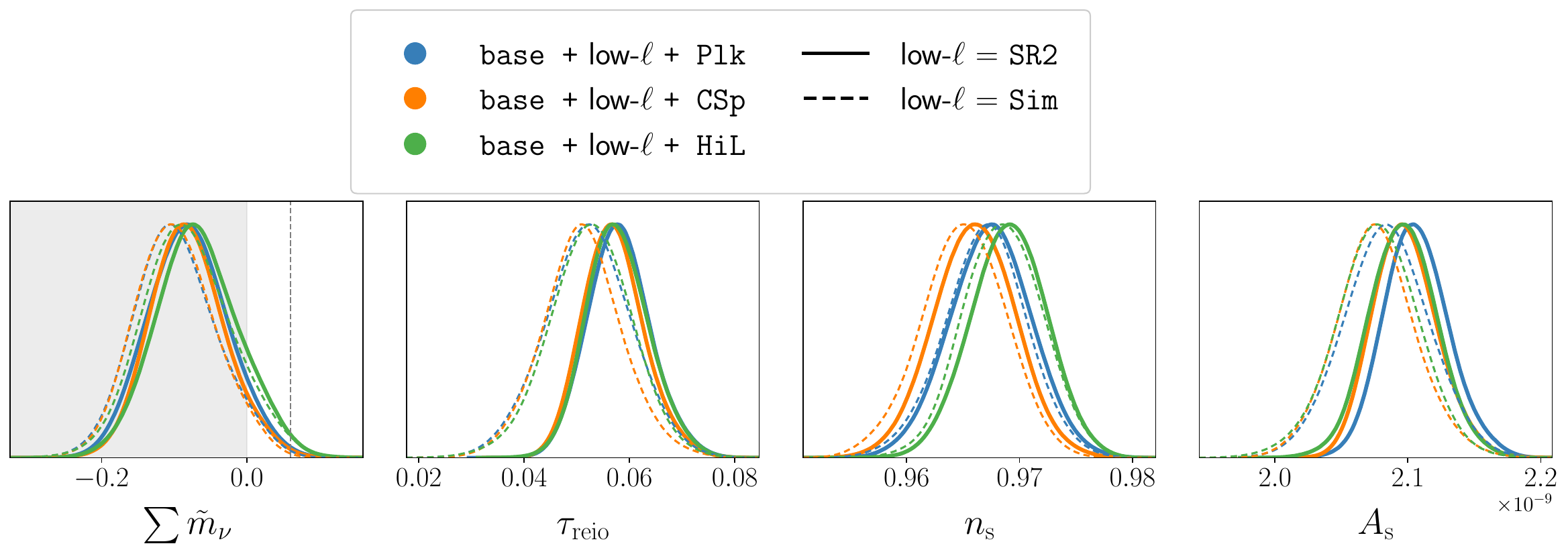}
    \caption{1D marginalized posterior distribution for $\summnu$, $\tau_\mathrm{reio}$, $n_s$, and $A_s$. All analyses include DESI DR2 BAO, \texttt{P-ACT} lensing, low-$\ell$ TT from \texttt{Commander} (\texttt{base}), and low-$\ell$ EE from \texttt{SRoll2}. Results are shown for \texttt{Plik} (blue), \texttt{CamSpec} (orange), and \texttt{HiLLiPoP} (green). The dashed vertical line marks the indicates the NO limit for the neutrino mass, $\sum m_\nu = 0.06~\mathrm{eV}$. The grey shaded region denotes the unphysical part of the parameter space.}
    \label{fig:high_l_marginal}
\end{figure*}

\begin{figure*}[t]
    \centering
    \includegraphics[width=\textwidth]{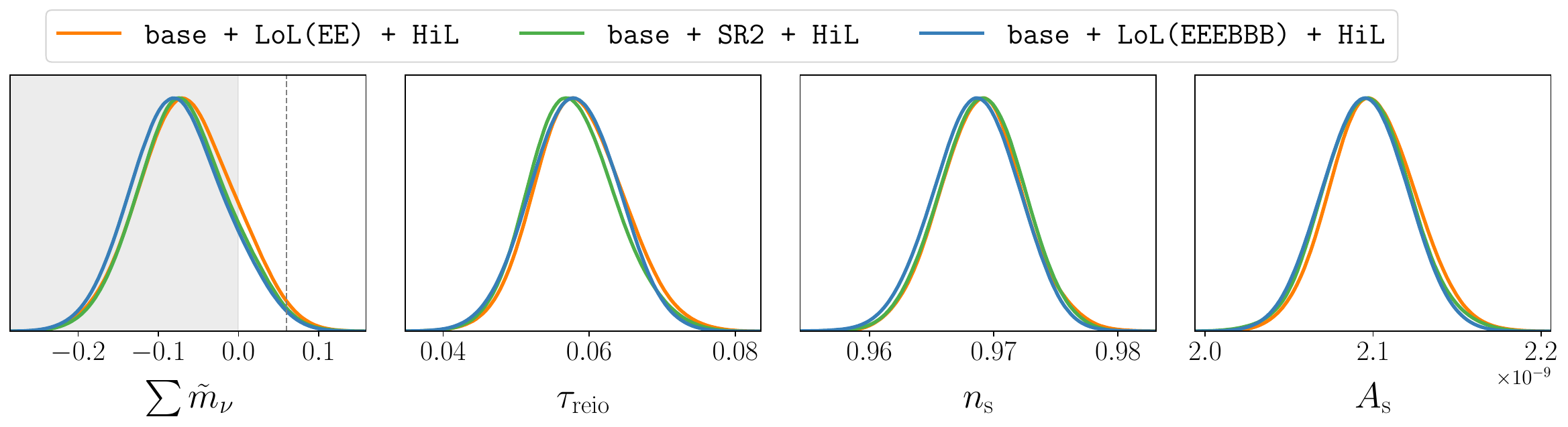}
    \caption{1D marginalized posterior distribution for $\sum \tilde{m}_\nu$, $\tau_\mathrm{reio}$, $n_s$, and $A_s$. All analyses include the \texttt{base} data sets (DESI DR2 BAO, \texttt{P-ACT} lensing, low-$\ell$ TT from \texttt{Commander}) and high-$\ell$ TTTEEE from \texttt{HiLLiPoP}. Results are shown for \texttt{LoLLiPoP (EE, BB, EB)} (blue), \texttt{LoLLiPoP (EE)} (orange), and \texttt{SRoll2} (green). The dashed vertical line indicates the NO limit for the neutrino mass, $\sum m_\nu = 0.06~\mathrm{eV}$. The grey shaded region denotes the unphysical part of the parameter space.}
    \label{fig:low_l_marginal}
\end{figure*}
\section{Full comparison among \textit{Planck} likelihoods}\label{ap:app-planck}

In this Appendix, we investigate how the inferred neutrino mass depends on the choice of CMB temperature and polarization likelihoods~\cite{ACT:2025fju, DESI:2025gwf}. First, we compare high-$\ell$ likelihoods, namely \texttt{Plik}, \texttt{CamSpec}, and \texttt{HiLLiPoP}, while keeping the low-$\ell$ EE data set fixed to \texttt{SRoll2}.
\texttt{SRoll2}~\cite{Pagano:2019tci, Delouis:2019bub} provides a more accurate treatment of instrumental systematics than the default \textit{Planck} PR3 low-$\ell$ EE likelihood (\texttt{SimAll}~\cite{Planck:2019nip}). We also compare these results with \texttt{LoLLiPoP}, an alternative large-scale polarization likelihood derived from the 2020 NPIPE (PR4) data release. \texttt{LoLLiPoP} includes EE, BB, and EB spectra~\cite{Tristram:2020wbi, Tristram:2021tvh}. In recent analyses~\cite{Naredo-Tuero:2024sgf, DESI:2025ejh}, the full \texttt{LoLLiPoP} likelihood—comprising EE, BB, and EB—is used. To ensure a fair comparison, we isolate the EE-only component of \texttt{LoLLiPoP}~\cite{Allali:2024aiv, Asorey:2025hgx} and examine how the inclusion of B-mode data influences our findings. Unless otherwise stated, all our results include the baseline data sets, \textit{i.e.} DESI BAO, \texttt{P-ACT} lensing data set and low-$\ell$ TT from \texttt{Commander} (\texttt{base}) introduced in Section~\ref{sec:data}. 

In Figure~\ref{fig:high_l} we show 2D contours highlighting key degeneracies between $\summnu$ and $\tau_\mathrm{reio}$, $n_s$, and $A_s$. For \texttt{Plik} (blue) and \texttt{CamSpec} (orange), the NO mass limit lies just outside the $2\sigma$ region, while for \texttt{HiLLiPoP} (green) it falls within $2\sigma$. Compatibility with $\sum m_\nu = 0.06~\mathrm{eV}$ is $2.6\sigma$, $2.5\sigma$, and $2.2\sigma$ for \texttt{Plik}, \texttt{CamSpec}, and \texttt{HiLLiPoP}, respectively.

The combination of data sets allows for visible degeneracies with $n_s$, $A_s$, and $\tau_\mathrm{reio}$. In terms of Pearson correlation coefficient (see main text for further explanation) \texttt{HiLLiPoP} shows slightly higher correlation between $\summnu$ and $n_s$ \textit{i.e.} $\mathcal{P}(\summnu$-$n_s) = 0.41$ with respect to \texttt{Plik} (0.38) and \texttt{CamSpec} (0.28). In the case of $\tau_\mathrm{reio}$, \texttt{HiLLiPoP} shows a smaller correlation $\mathcal{P}(\summnu$-$\tau_\mathrm{reio}) = 0.31$ with respect to the other two, where it stands at $\sim 0.36$.

As expected, the low-$\ell$ EE data set primarily sets the constraint on $\tau_\mathrm{reio}$, which remains consistent across all tested configurations.
Similarly, $A_s$ shows negligible variation among likelihoods.
More notable differences arise in $n_s$ and $\omega_\mathrm{c}$. The latter is directly affected by the increased neutrino mass, which contributes to the total energy density through $\omega_\nu \simeq \sum m_\nu / 93.14\,\mathrm{eV}$, resulting in a slight reduction in $\omega_\mathrm{c}$ (see Table~\ref{tab:part1} in Appendix~\ref{ap:app-additional}). Despite this, the total matter abundance $\Omega_\mathrm{m}$ remains stable and agrees with the DESI measurement~\cite{DESI:2025ejh}, although it is lower than typical CMB-only estimates~\cite{Planck:2018vyg}, consistent with the known effect of including BAO data.
Interestingly, we observe a positive degeneracy between $\summnu$ and $n_s$, in contrast to the negative correlation often seen in CMB-only studies~\cite{Archidiacono:2016lnv}. This degeneracy reversal can be explained as follows.
Primary CMB data sets are known to exhibit degeneracies between the total neutrino mass $\sum \tilde{m}_\nu$ and several cosmological parameters. These include background parameters -- such as $H_0$ and  $\omega_\mathrm{c}$ -- as well as parameters like tilt and amplitude of the primordial power spectrum, and $\tau_\mathrm{reio}$~\cite{Archidiacono:2016lnv}. The inclusion of BAO data, which offer precise distance measurements, is essential for constraining the expansion history of the universe. By tightly bounding parameters such as $H_0$ and $\omega_\mathrm{c}$, BAO observations significantly reduce the freedom to compensate for variations in $\sum \tilde{m}_\nu$ through adjustments in geometric quantities. As a result, the correlation between $\sum \tilde{m}_\nu$ and $n_s$ reverses direction compared to what is reported in Figure 2 of~\cite{Archidiacono:2016lnv}. 
We also find a positive degeneracy between $\summnu$ and $A_s$. To compensate for the suppression of the gravitational lensing due to increasing the neutrino masses, $A_s$ must increase.
In Figure~\ref{fig:high_l_marginal} we show the corresponding 1D marginalized posterior distributions for $\summnu$, $\tau_\mathrm{reio}$, $n_s$, and $A_s$. Solid lines correspond to \texttt{SRoll2}, while dashed lines show results with \texttt{SimAll} for low-$\ell$ EE. Compared to Table V of~\cite{DESI:2025ejh}, which used \texttt{SimAll} with \texttt{Plik} or \texttt{CamSpec}, the use \texttt{SRoll2} shifts the posterior distribution mean slightly toward positive neutrino masses without reducing constraining power. For the cases using \texttt{SRoll2+Plik} and \texttt{SRoll2+CamSpec}, we report posterior distribution means (best-fit) of $\sum \tilde{m}_\nu = -0.081^{+0.054}_{-0.057}$ ($-0.120$) eV and $\summnu = -0.080^{+0.046}_{-0.056}$ ($-0.116$) eV, respectively. 
\texttt{SRoll2 + HiLLiPoP} combination yields a mean (best-fit) estimate of $\sum \tilde{m}_\nu = -0.068^{+0.054}_{-0.061}$ ($-0.107$) eV. 

In Figure~\ref{fig:high_l_marginal}, we observe an upward shift in $n_s$ when using \texttt{HiLLiPoP}, compared to \texttt{Plik} or \texttt{CamSpec}. However, inferred value remains statistically consistent with \textit{Planck} 2018 result~\cite{Planck:2018vyg}, as it falls within the $1\sigma$ range. This trend is consistent with the ACT DR6 release~\cite{ACT:2025fju}, indicating that, for neutrino mass constraints, a primary degeneracy is with $n_s$.

In Figure~\ref{fig:low_l_marginal} we examine the impact of different low-$\ell$ EE likelihoods, showing the 1D marginalized posterior distributions for $\summnu$, $\tau_\mathrm{reio}$, $n_s$, and $A_s$. The green contours correspond to the reference configuration \texttt{SRoll2 + HiLLiPoP}. The orange contours represent the results using the EE-only version of the \texttt{LoLLiPoP} likelihood, while the blue contours correspond to the full \texttt{LoLLiPoP} configuration including EE, BB and EB data. We do not report any appreciable shift in the parameters inference.
The inferred mass of the neutrinos (best-fit) is $\summnu = -0.064^{+0.058}_{-0.062}(-0.111)\, \mathrm{eV}$ and  $\summnu = -0.075^{+0.055}_{-0.063}(-0.070)\, \mathrm{eV}$ for the EE-only \texttt{LoLLiPoP + HiLLiPoP} and the full \texttt{LoLLiPoP + HiLLiPoP} setup (with B-modes), respectively.

\section{Additional tables}\label{ap:app-additional}
In this Appendix, we present the complete set of tables listing all cosmological parameters for the main configurations analyzed in this work. The tables include the six standard \lcdm parameters: $\log(10^{10}A_s)$, $n_s$, $\theta_s$, $\omega_\mathrm{b}$, $\omega_\mathrm{c}$, and $\tau_\mathrm{reio}$, along with the effective total neutrino mass $\summnu$ (in eV), and the two lensing parameters, $A^{\phi\phi}_\mathrm{lens}$ and $A^{\mathrm{TTTEEE}}_\mathrm{lens}$.
The second section of each table lists derived parameters, including the Hubble constant $H_\mathrm{0}$, the matter fluctuation amplitude at $8 h^{-1} \mathrm{Mpc}$, \textit{i.e.}~$\sigma_8$, the reionization redshift $z_\mathrm{reio}$, the matter density parameter $\Omega_\mathrm{m}$, and the age of the universe in Gyr.
Each configuration includes the \texttt{base} data set, which comprises DESI DR2 BAO measurements, \texttt{P-ACT} lensing data, and low-$\ell$ temperature measurements from \texttt{Commander}. Table~\ref{tab:part1} compares the results obtained using the three high-$\ell$ TTTEEE likelihoods from \textit{Planck}—\texttt{Plik}, \texttt{CamSpec}, and \texttt{HiLLiPoP}—with the low-$\ell$ EE likelihood fixed to \texttt{SRoll2}. Table~\ref{tab:part2} uses the \texttt{SRoll2 + HiLLiPoP} CMB likelihood combined with the \texttt{base} configuration, here referred to as the \texttt{baseline}. It compares three cases: one where $A^{\mathrm{TTTEEE}}_\mathrm{lens}$ varies with $A^{\phi\phi}_\mathrm{lens}=1$, one where $A^{\phi\phi}_\mathrm{lens}$ varies with $A^{\mathrm{TTTEEE}}_\mathrm{lens}=1$, and a final case where both parameters are allowed to vary simultaneously.

\begin{table*}[htbp!]
    \caption{Summary of cosmological constraints obtained using different high-$\ell$ \textit{Planck} TTTEEE likelihoods—\texttt{Plik}, \texttt{CamSpec}, and \texttt{HiLLiPoP}—combined with low-$\ell$ EE polarization data from \texttt{SRoll2}. All configurations include the \texttt{base} data set, which consists of DESI DR2 BAO, \texttt{P-ACT} lensing, and low-$\ell$ TT data from \texttt{Commander}. We report the posterior distribution means and 68$\%$ C.L., with best-fit values shown in parentheses. The listed parameters include the six standard \lcdm parameters and the total neutrino mass $\summnu$, along with the derived parameters $H_0$, $\sigma_8$, $z_\mathrm{reio}$, $\Omega_\mathrm{m}$, and the age of the universe.} 
    \centering
    \begin{tabular}{l@{\hskip 14pt}c@{\hskip 27pt}c@{\hskip 27pt}c}
    \hline 
    \noalign{\vskip 0.5ex}
 & \texttt{base + SR2 + Plk} & \texttt{base + SR2 + CSp} & \texttt{base + SR2 + HiL} \\[0.5ex]
\hline 
\noalign{\vskip 0.5ex}
$\log(10^{10}A_\mathrm{s})$ & $3.048_{-0.012}^{+0.011}$\,(3.043) & $3.044_{-0.012}^{+0.011}$\,(3.039) & $3.043 \pm 0.012$\,(3.049) \\[0.5ex]
$n_\mathrm{s}$ & $0.9674_{-0.0036}^{+0.0037}$\,(0.9678) & $0.9660 \pm 0.0036$\,(0.9648) & $0.9690_{-0.0034}^{+0.0033}$\,(0.9699) \\[0.5ex]
$\theta_s$ & $1.04195 \pm 0.00027$\,(1.04186) & $1.04182 \pm 0.00023$\,(1.04181) & $1.04185_{-0.00022}^{+0.00023}$\,(1.04184) \\[0.5ex]
$\omega_{\mathrm{b}}$ & $0.02244_{-0.00014}^{+0.00013}$\,(0.02248) & $0.02227 \pm 0.00013$\,(0.02224) & $0.02229 \pm 0.00012$\,(0.02229) \\[0.5ex]
$\omega_{\mathrm{c}}$ & $0.11891_{-0.00087}^{+0.00078}$\,(0.11921) & $0.11870_{-0.00075}^{+0.00074}$\,(0.11914) & $0.11845_{-0.00087}^{+0.00079}$\,(0.11830) \\[0.5ex]
$\tau_\mathrm{reio}$ & $0.0580_{-0.0062}^{+0.0056}$\,(0.0565) & $0.0570_{-0.0063}^{+0.0052}$\,(0.0542) & $0.0578_{-0.0064}^{+0.0054}$\,(0.0600) \\[0.5ex]
$\sum \tilde m_\nu \,[\mathrm{eV}]$ & $-0.081_{-0.056}^{+0.054}$\,(-0.098) & $-0.081_{-0.055}^{+0.047}$\,(-0.111) & $-0.068_{-0.061}^{+0.054}$\,(-0.048) \\[0.5ex]
\hline 
\noalign{\vskip 0.5ex}
$H_0 \,[\mathrm{km}/\mathrm{s}/\mathrm{Mpc}]$ & $69.00 \pm 0.37$\,(69.04) & $68.88_{-0.39}^{+0.38}$\,(68.94) & $68.88_{-0.37}^{+0.36}$\,(68.73) \\[0.5ex]
$\sigma_8$ & $0.839 \pm 0.012$\,(0.842) & $0.837 \pm 0.011$\,(0.843) & $0.834_{-0.013}^{+0.012}$\,(0.831) \\[0.5ex]
$z_\mathrm{reio}$ & $7.99_{-0.58}^{+0.57}$\,(7.85) & $7.93_{-0.61}^{+0.53}$\,(7.66) & $8.00_{-0.61}^{+0.55}$\,(8.23) \\[0.5ex]
$\Omega_\mathrm{m}$ & $0.2948_{-0.0045}^{+0.0044}$\,(0.2947) & $0.2950_{-0.0045}^{+0.0047}$\,(0.2945) & $0.2948 \pm 0.0044$\,(0.2964) \\[0.5ex]
$\mathrm{Age}\,[\mathrm{Gyr}]$ & $13.718 \pm 0.025$\,(13.710) & $13.739_{-0.025}^{+0.026}$\,(13.730) & $13.741 \pm 0.024$\,(13.753) \\[0.5ex]
\hline 
\noalign{\vskip 0.5ex}
\label{tab:part1}
\end{tabular}
\end{table*}

\begin{table*}[htbp!]
    \caption{Summary of cosmological constraints obtained using the \texttt{SRoll2} low-$\ell$ polarization likelihood and the \texttt{HiLLiPoP} high-$\ell$ TTTEEE likelihood. All configurations also include DESI DR2 BAO, \texttt{P-ACT} lensing, and low-$\ell$ TT data from \texttt{Commander}. For simplicity, we refer to this combination as \texttt{baseline}.
    From left to right, the first column shows the case where $A^{\mathrm{TTTEEE}}_\mathrm{lens}$ varies while $A^{\phi\phi}_\mathrm{lens}=1$; the second column shows the case where $A^{\phi\phi}_\mathrm{lens}$ varies while $A^{\mathrm{TTTEEE}}_\mathrm{lens}=1$; and the last column shows the case where both lensing amplitude parameters are allowed to vary.
    We report the posterior distribution means and 68$\%$ C.L., with best-fit values shown in parentheses. The listed parameters include the six standard \lcdm parameters and the total neutrino mass $\summnu$,, along with the derived parameters $H_0$, $\sigma_8$, $z_\mathrm{reio}$, $\Omega_\mathrm{m}$, and the age of the universe.} 
    \centering
    \begin{tabular}{l@{\hskip 14pt}c@{\hskip 27pt}c@{\hskip 27pt}c}
    \hline 
    \noalign{\vskip 0.5ex}
 & \texttt{baseline} - $A_\mathrm{lens}^{\mathrm{TTTEEE}}$ & \texttt{baseline} - $A_\mathrm{lens}^{\phi\phi}$ & \texttt{baseline} - $A_\mathrm{lens}^{\mathrm{TTTEEE}+\phi\phi}$ \\[0.5ex]
\hline 
\noalign{\vskip 0.5ex}
$\log(10^{10}A_\mathrm{s})$ & $3.042_{-0.013}^{+0.011}$\,(3.051) & $3.039 \pm 0.014$\,(3.054) & $3.036_{-0.015}^{+0.013}$\,(3.029) \\[0.5ex]
$n_\mathrm{s}$ & $0.9686_{-0.0038}^{+0.0039}$\,(0.9698) & $0.9698_{-0.0036}^{+0.0035}$\,(0.9691) & $0.9706_{-0.0040}^{+0.0043}$\,(0.9715) \\[0.5ex]
$\theta_s$ & $1.04184_{-0.00025}^{+0.00024}$\,(1.04181) & $1.04189_{-0.00025}^{+0.00024}$\,(1.04184) & $1.04189 \pm 0.00024$\,(1.04178) \\[0.5ex]
$\omega_{\mathrm{b}}$ & $0.02227 \pm 0.00013$\,(0.02236) & $0.02230 \pm 0.00012$\,(0.02233) & $0.02233_{-0.00013}^{+0.00015}$\,(0.02227) \\[0.5ex]
$\omega_{\mathrm{c}}$ & $0.11852_{-0.00086}^{+0.00090}$\,(0.11836) & $0.11809_{-0.00090}^{+0.00087}$\,(0.11806) & $0.1178_{-0.0013}^{+0.0011}$\,(0.1177) \\[0.5ex]
$\tau_\mathrm{reio}$ & $0.0582_{-0.0063}^{+0.0055}$\,(0.0632) & $0.0580_{-0.0065}^{+0.0057}$\,(0.0622) & $0.0572_{-0.0063}^{+0.0052}$\,(0.0538) \\[0.5ex]
$\sum \tilde m_\nu \,[\mathrm{eV}]$ & $-0.071_{-0.068}^{+0.059}$\,(-0.084) & $-0.038_{-0.068}^{+0.064}$\,(-0.024) & $-0.018_{-0.089}^{+0.085}$\,(-0.001) \\[0.5ex]
$A_\mathrm{lens}^{\phi\phi}$ & - & $1.028_{-0.037}^{+0.034}$\,(1.008) & $1.043_{-0.050}^{+0.049}$\,(1.057) \\[0.5ex]
$A_\mathrm{lens}^{\mathrm{TTTEEE}}$ & $0.987 \pm 0.043$\,(0.989) & - & $1.026 \pm 0.062$\,(1.044) \\[0.5ex]
\hline 
\noalign{\vskip 0.5ex}
$H_0 \,[\mathrm{km}/\mathrm{s}/\mathrm{Mpc}]$ & $68.9 \pm 0.4$\,(69.1) & $68.8 \pm 0.4$\,(68.7) & $68.8 \pm 0.4$\,(68.6) \\[0.5ex]
$\sigma_8$ & $0.835 \pm 0.014$\,(0.840) & $0.825_{-0.016}^{+0.015}$\,(0.827) & $0.819 \pm 0.022$\,(0.813) \\[0.5ex]
$z_\mathrm{reio}$ & $8.04_{-0.61}^{+0.56}$\,(8.51) & $8.01_{-0.60}^{+0.59}$\,(8.43) & $7.92_{-0.61}^{+0.53}$\,(7.60) \\[0.5ex]
$\Omega_\mathrm{m}$ & $0.2949 \pm 0.0047$\,(0.2926) & $0.2955 \pm 0.0047$\,(0.2973) & $0.2959 \pm 0.0046$\,(0.2975) \\[0.5ex]
$\mathrm{Age}\,[\mathrm{Gyr}]$ & $13.741_{-0.027}^{+0.026}$\,(13.728) & $13.750 \pm 0.026$\,(13.758) & $13.754 \pm 0.028$\,(13.772) \\[0.5ex]
\hline 
\noalign{\vskip 0.5ex}
\label{tab:part2}
\end{tabular}
\end{table*}

\end{document}